\documentclass[fleqn,usenatbib]{mnras}
\usepackage{newtxtext,newtxmath}
\usepackage{orcidlink}

\usepackage[T1]{fontenc}

\DeclareRobustCommand{\VAN}[3]{#2}
\let\VANthebibliography\thebibliography
\def\thebibliography{\DeclareRobustCommand{\VAN}[3]{##3}\VANthebibliography}

\usepackage{graphicx}	
\usepackage{amsmath}	
\usepackage{float}
\usepackage{multirow}
\usepackage{caption}
\usepackage{subcaption}
\usepackage{xcolor}
\usepackage{physics}
\usepackage{enumitem}

\newcommand{\Z}[1]{_{\lower2pt\hbox{$\scriptstyle#1$}}}
\newcommand{\Ns}[1]{_{\lower2pt\hbox{$\scriptstyle\rm#1$}}}

\newcommand{\LCDM}{$\Lambda$CDM }
\newcommand{\hm}{\,h^{-1}\hbox{Mpc}}
\newcommand{\ab}{\bar a}
\newcommand{\goesas}{\,{\mathop{\sim}\limits}\,}
\def\Z#1{_{\lower2pt\hbox{$\scriptstyle#1$}}}
\definecolor{DRed}{rgb}{0.71,0.14,0.07}

\definecolor{CBlue}{rgb}{0.0,0.7,0.8}

\definecolor{CGreen}{rgb}{0.0,0.5,0.0}

\title[Revisiting lens mass models in cosmology]{Revisiting the effect of lens mass models in cosmological applications of strong gravitational lensing}

\author[C. Harvey-Hawes et al.]{
Christopher Harvey-Hawes\orcidlink{0009-0002-6751-2695}\thanks{E-mail: christopher.harvey-hawes@pg.canterbury.ac.nz },
David L. Wiltshire\orcidlink{0000-0003-1992-6682}
\\
School of Physical \& Chemical Sciences, University of Canterbury, Private Bag 4800, Christchurch 8041, New Zealand\\
}

\date{Accepted XXX. Received YYY; in original form ZZZ}

\pubyear{2024}

\begin{document}
\label{firstpage}
\pagerange{\pageref{firstpage}--\pageref{lastpage}}
\maketitle

\begin{abstract}
Strong gravitational lens system catalogues are typically used to constrain a combination of cosmological and empirical power-law lens mass model parameters, often introducing additional empirical parameters and constraints from high resolution imagery. We investigate these lens models using Bayesian methods through a novel alternative that treats spatial curvature via the non-FLRW timescape cosmology. We apply Markov Chain Monte Carlo methods using the catalogue of 161 lens systems of \citet{chenAssessingEffectLens2019} in order to constrain both lens and cosmological parameters for: (i) the standard \LCDM model with zero spatial curvature; and (ii) the timescape model. We then generate large mock data sets to further investigate the choice of cosmology on fitting simple power-law lens models. In agreement with previous results, we find that in combination with single isothermal sphere parameters, models with zero FLRW spatial curvature fit better as the free parameter approaches an unphysical empty universe, $\Omega_{\mathrm M0}\to0$. By contrast, the timescape cosmology is found to prefer parameter values in which its cosmological parameter, the present void fraction, is driven to $f_{\mathrm v0}\to0.73$ and closely matches values that best fit independent cosmological data sets: supernovae Ia distances and the cosmic microwave background. This conclusion holds for a large range of seed values $f_{\mathrm v0}\in\{0.1,0.9\}$, and for timescape fits to both timescape and FLRW mocks. Regardless of cosmology, unphysical estimates of the distance ratios given from power-law lens models result in poor goodness of fit. With larger datasets soon available, separation of cosmology and lens models must be addressed.
\end{abstract}


\parskip=5pt plus.2pt minus.1pt
\section{Introduction}
 Strong gravitational lensing (SGL) occurs when the light path from a distant source is warped around an intermediary lens body on its way to an observer, producing several images, arcs or rings. Over the last couple of decades, strong gravitational lenses have been observed with an increased number. The current largest compilation of all strong gravitational lenses which can be used viably for constraining lens properties and cosmological parameters consists of 161 systems \citep{chenAssessingEffectLens2019}. The number of such strong gravitational lenses is set to increase by orders of magnitude over the coming decade: the Rubin Observatory Large Synoptic Survey Telescope (LSST) and Euclid are projected to observe $1.2\times10^6$ and $1.7\times10^6$ galaxy-galaxy scale lenses respectively \citep{collettPOPULATIONGALAXYGALAXY2015}. Since the standard cosmology faces increasing challenges \citep{buchertObservationalChallengesStandard2016,peeblesAnomaliesPhysicalCosmology2022,aluriObservableUniverseConsistent2023} it is important to understand the interplay of lens models and cosmological models. In this paper, we will test such assumptions using the existing \citet{chenAssessingEffectLens2019} catalogue.

It is possible to constrain the properties of the lens galaxies given an assumed background cosmology, typically using the standard spatially flat $\Lambda$ Cold Dark Matter (\LCDM) model with Planck value parameters \citep{planckcollaborationPlanck2018Results2020}. Alternatively, one can constrain the parameters of the assumed cosmological model given a particular choice of lens mass density profile. Attempts to simultaneously fit both the lens density profile and cosmological parameters have been conducted \citep{caoCosmologyStrongLensing2015,chenAssessingEffectLens2019}, inevitably running into degeneracies that require further observational data to resolve.

There are several different methods to perform statistical cosmological analysis with strong gravitational lenses. Time-delay cosmography as pioneered by \citet{refsdalPossibilityDeterminingHubble1964} is the most sensitive to cosmological parameters and deals with each system individually rather than trying to deal with an `average' ideal lens model. This method is particularly useful in determining the Hubble constant, $H_0\equiv H(t_0)$, as the measured time-delay between multiply imaged events can be used to determine their different path lengths, and thereby a value for the Hubble expansion \citep{wongH0LiCOWXIIIMeasurement2020,millonTDCOSMOExplorationSystematic2020}. Despite time-delay cosmography being a powerful technique, only a small fraction of the observed lensing systems have well-defined time-delays, thus limiting the potential application of this approach. The vast increase in available lens systems from the next generation of telescopes is now set to overcome this hurdle.

To date most observed galaxy-scale lensing systems involve distant quasars sources. As compared to observations of smaller numbers of lensed SneIa, the uncertainties in time-delay measurements of quasars are significantly larger. Given the paucity of data available for accurate time-delay cosmography, alternate approaches should be considered to make full use of the observed lens systems. 

This paper will focus in particular on the distance ratio
\begin{equation} \label{eq:Dratiosumrule}
    \frac{d_{ ls}}{d_{ s}} = \sqrt{1-k\, d_{ l}^2} - \frac{d_{ l}}{d_{ s}} \,\sqrt{1-k\, d_{ s}^2} \quad ,
\end{equation}
introduced in the distance sum rule test of \citet{rasanenNewTestFLRW2015}. Here $d_s$, $d_l$ and $d_{ls}$ denote the dimensionless comoving distances, $d(z_l,z_s)$, from the source to observer, lens to observer and source to lens respectively. The comoving distance is related to the angular diameter distances, $D_A$, and dimensionless luminosity distance, $D_L$ by 
\begin{align}
    d &=\frac{H_0}{c}D_A(1+z)=\frac{H_0}{c}\frac{D_L}{1+z} \\
     &= \frac{1}{\sqrt{\Omega_{\rm k0}}} \sinh{ \sqrt{\Omega_{\rm k0}} \int_1^{1+z}\frac{{\rm d}x}{\sqrt{\Omega_{\Lambda0} +\Omega_{\rm k0}\, x^2 + \Omega_{\rm M0} \, x^3 }} }, \label{eq:FLRW_full}
\end{align}
where $\Omega_{\rm k0} = -kc^2/H_0^2 = 1 - \Omega_{\rm M0} - \Omega_{\Lambda0}$, and $\Omega_{\rm M0}$ and $\Omega_{\Lambda0}$ are the present epoch matter and cosmological constant density parameters. In this paper, we will focus on spatially flat FLRW models where $\Omega_{\rm k0} =0$ and thus \eqref{eq:FLRW_full} becomes
\begin{equation}
    d = \int_1^{1+z}\frac{{\rm d}x}{\sqrt{(1 - \Omega_{\rm M0}) + \Omega_{\rm M0} \, x^3 }} \,. \label{eq:FLRW_d}
\end{equation}
and the distance ratio is given as
\begin{equation}
   \mathcal{D} \equiv \frac{d_{ls}}{d_s} = \frac{d_s - d_l}{d_s} \,. \label{eq:FLRW_dratio}
\end{equation}

The relation \eqref{eq:Dratiosumrule} provides an exact expression for the ratio of two observed angular distances in the case of a FLRW metric with constant spatial curvature $k$.
It is then typically used as a self-consistency check of the standard FLRW cosmology, or to constrain the global spatial curvature of some assumed FLRW universe \citep{liaoTestFLRWMetric2017,qiStronglyGravitationallyLensed2019}. However, any cosmological model -- whether FLRW or not -- can be substituted on the r.h.s. of \eqref{eq:Dratiosumrule} provided the model has a suitable definition of the distance ratio in terms of its underlying parameters. 

For a specific lensing system, it is also possible to compute its respective distance ratio purely from a given lens model and its corresponding observables and parameter values, independent of cosmology. Therefore, we can compare values for the distance ratio derived from cosmological angular diameter distance measurements to values determined from an assumed lens model, thus, allowing for constraints on both lens and cosmological parameters. This can be extended to a comparison of different cosmological models through Bayesian inference \citep{cardoneCosmologicalParametersLenses2016}.
Such Bayesian analysis might involve two FLRW models with different priors on cosmological parameters, or alternatively an FLRW model compared to a non-FLRW model such as the timescape \citep{wiltshireAverageObservationalQuantities2009}.

The distance ratio is independent of the Hubble parameter, and therefore sidesteps the issue of its current epoch value $H_0=H(t_0)$. This is a major advantages of this technique, as the Hubble tension continues to pose a significant challenge to the \LCDM model \citep{riessCrowdedNoMore2023}. Furthermore, since SGL bypasses the cosmic distance ladder, it directly allows us to infer cosmological parameters without relying on other astronomical distance measures and the systematics they depend upon.

While the distance ratio has been used to test the self-consistency of FLRW models, to date it has not been applied to non-FLRW metrics.
This paper will fill that gap by revisiting the analysis of \citet{chenAssessingEffectLens2019} and extending it to include the timescape model.
We will perform a Bayesian comparison between the spatially flat FLRW model and the timescape model, and determine the corresponding best fit cosmological parameters. In addition, we investigate three different parametrisations of the lens galaxies mass profile and their effect on cosmological fits. Strong gravitational lensing is heavily dependent on the choice of lens model, as well as the background cosmology the systems are embedded within. Therefore, it is vital that lens galaxies are well modelled in advance of new observations by the next generation of telescopes.   

An overview of the paper is as follows. In Section \ref{sec:Timescape Overview}, we will briefly review the timescape model. In Section \ref{sec:Catalogue and Lens Models} we will discuss the different lens models used in this paper, as well as provide a description of the catalogue of lensing systems, identifying the uncertainties involved and model biases. Section \ref{sec:Results} outlines the methodology for determining cosmological parameters within a given lens model and gives the results of the fitting procedure. Section \ref{sec:Simulations} presents the results of simulations of mock data generated from the catalogue data with randomly added noise. In section \ref{sec:Conclusions} we discuss our results and the conclusions that can be drawn.

\section{Timescape Model Overview}
\label{sec:Timescape Overview}
Standard \LCDM cosmology, and indeed all FLRW models, are based upon the cosmological principle that on average the universe is both spatially homogeneous and isotropic. While some notion of spatial homogeneity and isotropy certainly holds for whole sky averages of very distant objects, the universe is inhomogeneous on scales $\lesssim$100$\hm$ \citep{eisensteinDetectionBaryonAcoustic2005,Sylos_Labini_2009,Scrimgeour_2012} where it is dominated by a cosmic web of over-dense filaments and under-dense voids. The standard cosmology has been incredibly successful.  
Nonetheless, challenges to the $\Lambda$CDM model remain unresolved despite of the growth of available data and increased observational precision (\citet{buchertObservationalChallengesStandard2016, aluriObservableUniverseConsistent2023, peeblesAnomaliesPhysicalCosmology2022}).

The timescape model assumes that cosmology should not invoke a single global reference background with a unique split of space and time
\citep{wiltshireCosmicClocks2007a,wiltshireExactSolution2007b,wiltshireTimeTimescapeEinstein2009a}. To model the universe effectively, we have to account for {\em backreaction of inhomogeneities}, namely deviations from average FLRW evolution at the present epoch. This is due to structure formation of filaments and voids and the resulting cosmic web, which grows in complexity in the late universe.
Timescape makes use of Buchert spatial averages \citep{buchertAveragePropertiesInhomogeneous2001} to provide the average evolution of the Einstein equations with backreaction.

The interpretation of Buchert's formalism has been much debated; critics pointed out that as a fraction of the averaged energy density, a term Buchert denotes the ``kinematical backreaction'' $\bar\Omega_{\cal Q}$, should be small \citep{IshibashiWald2006}. A universal feature of all viable backreaction proposals is that, unlike FLRW, average spatial curvature does not scale as a simple power of the average cosmic scale factor $\ab$, $\bar\Omega_K\ne -k/(\ab^2\bar H^2)$, where $k$ is constant and $\bar H$ an average expansion rate.
In the timescape model, in particular, $\bar\Omega_K\propto f_{\rm v}^{1/3}/(\ab^2\bar H^2)$, where the variable {\em void volume fraction}, $f_{\rm v}\to 0$ at early times but is significant in the late epoch universe. This means that conclusions about spatial curvature derived from CMB data in FLRW models are not relevant for backreaction models: when fit to the angular diameter distance of the CMB (with $\Lambda=0$) it is curvature, $\bar\Omega_K$, that dominates in the late epoch universe, \citep[Fig.~1]{DuleyTimescapeRadiation2013}. Furthermore, while the average energy-density parameters for matter, radiation, curvature and kinematical backreaction satisfy a simple sum rule
\begin{equation}
\bar\Omega_M+\bar\Omega_R+\bar\Omega_K+\bar\Omega_{\cal Q}=1\,,\label{sumrule}
\end{equation}
the quantities appearing in \eqref{sumrule} are {\em statistical} volume average quantities, not directly related to {\em local} observables. Consequently, conclusions about apparent cosmic acceleration do not follow simply from determining the magnitude of $\bar\Omega_{\cal Q}$. The timescape ansatz fixes the relationship of these quantities to local observables, resulting in a viable phenomenology with $\bar\Omega_K\goesas0.86$, $\bar\Omega_M\goesas0.17$, $\bar\Omega_{\cal Q}\goesas-0.03$ at present, $z=0$.

Since the local geometry varies dramatically depending on the position of an observer in the cosmic web, local geometry is generally very different to the global volume average. Local structure not only affects the average evolution of the Einstein equations, but also the calibration of distance and time measures relative to the volume average\footnote{A volume average observer in the timescape model represents an observer whose local spatial curvature coincides with that of the largest spatial averages. Such an observer is necessarily at an average position by volume - in a void - and systematically different from observers in bound structures. A typical location by volume is not a typical location by mass.} observer. Timescape allows for variation of the calibration of regional clocks relative to one another due to the gravitational energy cost of the gradients in spatial curvature between filaments and voids. Ideal observers, who see a close to isotropic CMB, in different local environments would age differently, hence the name timescape.

The accelerated expansion of the universe is now an apparent effect derived from fitting an FLRW model with constant spatial curvature and global cosmic time to an inhomogeneous non-FLRW universe. An ideal void observer not gravitationally bound to any structure will not infer a cosmic acceleration at all, whereas an observer in a bound structure such as galaxy will use a different time parameter and will infer an accelerated expansion at late cosmic epochs. The timescape model can eliminate the need for dark energy entirely and does so without introducing new ad hoc scalar fields or modifications to the gravitational action; rather it revisits averaging procedures that build upon Einstein's general relativity. 

Since the $\Lambda$CDM has been so successful in diverse cosmological tests, any phenomenologically viable model must yield similar predictions, particularly with regard to large scale averages for distances beyond the scale of inhomogeneity. For over a decade, timescape has consistently given fits to type Ia supernovae which are essentially statistically indistinguishable from \LCDM by Bayesian comparison \citep{Leith_2007,smaleSupernovaTestsTimescape2011,damApparentCosmicAcceleration2017}. The most recent analysis of the Pantheon+ catalogue shows, furthermore, that timescape predictions for cosmic expansion below $\goesas100\hm$ scales are consistent with observations in a manner which may provide a self-consistent resolution of the Hubble tension \citep{LaneCosmologicalFoundations2023}. The difference between the distance--redshift relations of the timescape model and those of {\em some} \LCDM model with fixed $\Omega_{M0}$, $\Omega_{\Lambda0}$, is only $1$--$3$\% over a small range of redshifts, but timescape effectively interpolates between \LCDM models with different parameters over larger redshift ranges -- which is why it can resolve the Hubble tension in a natural fashion. Of course, this must be tested with completely independent cosmological tests.

Among the many different definition of distance in the timescape model, here we focus on the matter dominated era, where a simple `tracking solution' is found to apply. 
Instead of the parameters $\Omega_{\rm M0}$ in a spatially flat FLRW model, distances are instead parameterised in terms of the present epoch void fraction $f_{\rm v0}$. The {\em effective comoving distances} directly comparable to those in the distance sum rule \eqref{eq:Dratiosumrule} are\footnote{Care must be taken with choices of units. With an explicit factor $c$, $b$ and $\mathcal{F}$ must have dimensions of $t$ and $t^{1/3}$ respectively, which is different to a convention often assumed for timescape \citep{Wiltshire_2014_cosmic}.}
\begin{equation}
    \begin{cases}
    d_s =  H_0 t_s^{2/3} [\mathcal{F}(t_0) - \mathcal{F}(t_s)] (1 + z_s) \\
    d_l = H_0 t_l^{2/3} [\mathcal{F}(t_0) - \mathcal{F}(t_l)] (1 + z_l)\\
    d_{ls} = H_0 t_s^{2/3} [\mathcal{F}(t_l) - \mathcal{F}(t_s)] (1 + z_s)  
    \end{cases}
    \label{eq:Timescape distances}
\end{equation}
where
\begin{multline}
    \mathcal{F}(t) = 2\,t^{1/3} + \frac{b^{1/3}}{6} \ln \left( \frac{(t^{1/3} + b^{1/3})^2}{t^{2/3} - b^{1/3}\,t^{1/3} + b^{2/3} } \right) \\ + \frac{b^{1/3}}{\sqrt{3}}\arctan \left( \frac{2t^{1/3} - b^{1/3}}{\sqrt{3}\, b^{1/3}} \right) \, ,
\end{multline}
\begin{equation}
    b = \frac{2\,(1-f_{\rm v0})(2+f_{\rm v0})}{9\,f_{\rm v0}\,\Bar{H}_0} \, ,
    \label{eq:b-Timescape}
\end{equation}
and $H_0$ and $\Bar H_0$ are the {\em dressed} and {\em bare} Hubble constants respectively.\footnote{The bare or volume average Hubble parameter is defined by the average volume expansion - not a global scale factor - and uses a time parameter relative to an idealised volume average observer. This time parameter differs systematically from that of observers in bound systems. The bare Hubble parameter is found to be related to the dressed Hubble parameter by
\begin{equation*}
H(t) = \frac{[4f_{\rm v}^2(t) + f_{\rm v}(t) +4] \Bar{H}(t)}{2(2+f_{\rm v}(t))} \, .
\end{equation*}
and $f_{\rm v0}\equiv f_{\rm v}(t_0)$ etc denote present epoch values.}
The Buchert volume averaged time parameter $t$ has been interpreted in different ways in different backreaction models. For the timescape tracker solution, observers in gravitationally bound systems measure a time parameter related to volume-average time according to
\begin{equation}
\tau=\frac{2}{3}t+\frac{2(1-f_{\rm v0})(2+f_{\rm v0})}{27f_{\rm v0}\Bar{H}_0}\,\ln{\left[1+\frac{9f_{\rm v0}\Bar{H}_0}{2(1-f_{\rm v0})(2+f_{\rm v0})}\right]} \,.
\end{equation}
The volume-average time parameter is related to our observed redshift by
\begin{equation}
    z + 1 = \frac{2^{4/3}\,t^{1/3}\,(t+b)}{f_{\rm v0}^{1/3}\,\Bar{H}_0 \,t \,(2t+3b)^{4/3}} \, ,
    \label{eq:z+1}
\end{equation}
 \citep{larenaTestingBackreactionEffects2009}. The distance ratio is then given as
\begin{equation}\label{eq:Timescape_ratio}
    \mathcal{D}^{\text{Time}} \equiv \frac{d_{ls}}{d_s} = \frac{\mathcal{F}(t_l) - \mathcal{F}(t_s)}{\mathcal{F}(t_0)-\mathcal{F}(t_s)} \,.
\end{equation}
Further details on the motivation and derivation of these equations is given by \citet{wiltshireAverageObservationalQuantities2009,Wiltshire_2014_cosmic}. It is now simple to make comparisons between timescape and spatially flat FLRW models, as they have the same number of free parameters: $f_{v0}$ and $\Omega_{M0}$ are direct analogues due to their effect on distance measurements.Any dependence on the dressed or bare Hubble constant is cancelled due to the nature of the distance ratio, hence there is only one defining cosmological parameter for each model.

\section{Catalogue, Observables and Lens Models}
\label{sec:Catalogue and Lens Models}
\subsection{Catalogue Data}
The data set used in this investigation is taken from the catalogue of 161 strongly lensed systems from \citet{chenAssessingEffectLens2019}. It is the largest catalogue compiled to date, and is itself built upon a previous catalogue of 118 systems from \citet{caoCosmologyStrongLensing2015}. This catalogue is a compilation of multiple surveys consisting of the LSD \citep{treuInternalStructureFormation2002,treuMassiveDarkmatterHalos2004,koopmansStellarVelocityDispersion2002,koopmansStructureDynamicsLuminous2003}, SL2S \citep{ruffSL2SGalaxyscaleLens2011,sonnenfeldSL2SGalaxyscaleLens2013,sonnenfeldSL2SGalaxyscaleLens2013a,sonnenfeldSL2SGalaxyscaleLens2015}, SLACS \citep{boltonSloanLensACS2008,augerSloanLensACS2009,augerSloanLensACS2010}, S4TM \citep{shuSloanLensACS2017,shuSLOANLENSACS2015}, BELLS \citep{brownsteinBOSSEmissionLineLens2012} and BELLS Gallery surveys \citep{shuBOSSEmissionLineLens2016,shuBOSSEmissionLineLens2016a}. 
From these surveys, the following observables are listed: spectroscopic lens and source redshift $z_l$ and $z_s$, Einstein radius $\theta_E$, half-light radius of the lens galaxy $\theta_\text{eff}$, aperture radius $\theta_\text{ap}$ and the central velocity dispersion of the lens galaxy $\sigma_\text{ap}$. Spectroscopic measurements are provided by Sloan Digital Sky Survey (SDSS),  W. M. Keck-II Telescope and Baryon Oscillation Spectroscopic Survey (BOSS) spectroscopic instruments depending on survey. Follow-up imagery from the Hubble Space Telescope (HST) Advanced Camera for Surveys is then used for the determination of $\theta_E$ and $\theta_\text{eff}$. 

Each of these surveys has differing instrumental parameters and thus has different uncertainties in the measurement of the central velocity dispersions\footnote{This is obtained by converting the redshift dispersion from spectroscopic measurements around the galaxy using the radial special relativistic formula $ z = \sqrt{(c + v)/(c-v)}-1 $.} of the lens galaxies, which contributes the largest portion of the uncertainty in this method. Velocity dispersions of lens galaxies are measured spectroscopically within a given aperture radius $\theta_{\rm ap}$. However, the shape and size of the aperture used in each survey differs and thus an aperture correction formula is applied so that velocity dispersions are determined as if they were found with a normalized typical circular aperture $\sigma_0$ \citep{jorgensenSpectroscopyS0Galaxies1995}.
To transform rectangular apertures into an equivalent circular aperture for comparison, the following formula is used,
\begin{equation}
    \theta_{\rm ap} \approx 1.025 \sqrt{\frac{\theta_x \theta_y}{\pi}} \, ,
\end{equation}
where $\theta_x$ and $\theta_y$ are the width and height of the rectangular aperture respectively. Since aperture radii vary even within the class of circular apertures, the following normalization is also applied along the line of sight,
\begin{equation} \label{eq:sigma0}
    \sigma_\parallel \equiv \sigma_0 = \sigma_{\text{ap}} \left( \frac{\theta_{\text{eff}}}{2\theta_{\text{ap}}} \right)^\eta \, ,
\end{equation}
where $\theta_{\text{eff}}$ corresponds to the half-light radius of the lens galaxy as $\theta_{\text{eff}} = R_{\text{eff}} / D_l$. The value of $\eta$ is taken from \citet{cappellariSAURONProjectIV2006} of $\eta = -0.066 \pm 0.035$  is found empirically via fitting to individual galaxy profiles and is in agreement with other values determined by \citet{jorgensenSpectroscopyS0Galaxies1995} and \citet{mehlertSpatiallyResolvedSpectroscopy2003} of $\eta = -0.04$ and $\eta = -0.06$ respectively.

The uncertainty in $\eta$ is the greatest source of uncertainty in the distance ratio inferred from gravitational lensing.
In addition, \eqref{eq:sigma0} contains a statistical error in $\sigma_{\text{ap}}$ as quoted in the surveys, and a further systematic error source. The latter is understood as the effect of interfering matter along the line of sight and whilst the surveys are restricted to isolated systems, the sytematic error is estimated at 3\% in all cases \citep{bernardiEarlyTypeGalaxiesSloan2003}. The combined total error budget is then given by, 
\begin{equation} \label{eq:uncert}
    \Delta \sigma_0^{\text{tot}} = \sqrt{(\Delta \sigma_0^{\eta})^2 + (\Delta \sigma_0^{\text{sys}})^2 + (\Delta \sigma_0^{\text{stat}})^2} \, .
\end{equation}
No uncertainty in the Einstein radius or associated redshifts is quoted within the catalogue data set. Thus, it is assumed these uncertainties are included within the systematic uncertainty of $\sigma_0$, although this is most likely an underestimation of the errors present in the data. A uniform error could be applied across the data set to $\theta_E$, but is omitted to stay consistent with the work of Chen et al. We find that changing the magnitude of the errors present in our parameters does not significantly effect our conclusions throughout this paper. 

Further cuts were also applied in compiling the catalogue. Not only did systems have to be isolated and have no significant substructure, but lens galaxies must also be early-type galaxies, either elliptical (E) or lenticular (S0).
\subsection{Lens Models}
When modelling lens mass distributions individually for use in time-delay cosmography, one of two descriptions are used: (i) power-law models that take into account both baryonic and dark matter simultaneously; and (ii) a combination of a luminous baryonic mass profile with a Navarro-Frank-White dark matter halo \citep{navarroStructureColdDark1996,wagnerSelfgravitatingDarkMatter2020}. Constraints on the value of $H_0$ from either class of model are very similar and statistically consistent with one another \citep{millonTDCOSMOExplorationSystematic2020}. However, to model a statistically average lens galaxy within large catalogues, more general power-law models are used \citep{koopmansGravitationalLensingStellar2006}.

In this section, we review the three power-law models applied to the \citep{chenAssessingEffectLens2019} catalogue, commonly used to model the matter distribution of elliptical lens galaxies: Extended Power-Law (EPL), Spherical Power-Law (SPL) and Singular Isothermal Sphere (SIS). 

All three lens models discussed in this paper rely upon the assumption that while galaxies may differ in shape, when taking averages the dominant component of the matter density is spherically symmetric. However, the angular structure of a lens galaxies mass distribution is very significant to the image separation produced. In fact, quadruply imaged sources are not possible with spherically symmetric mass distributions. Since the luminous matter in individual galaxies is not spherically symmetric, whether the angular component of all matter in each lens galaxy can be averaged out in this fashion is an assumption which can be tested in future.

Solving the radial Jeans equation, following the procedure given in Appendix~\ref{EPL Appendix}, the distance ratio for the most general Extended Power Law (EPL) model is found to be
\begin{equation} \label{eq:cleanmess}
    \mathcal{D}^{\text{obs}} \equiv \frac{d_{ls}}{d_s} = \frac{c^2 \theta_E}{2\sqrt{\pi} \sigma_0^2} \left( \frac{\theta_{\text{eff}}}{2\theta_E} \right)^{2-\gamma} F(\gamma,\delta,\beta) \, ,
\end{equation}
where $\theta_E$ is the Einstein radius of the system. $F(\gamma,\delta,\beta)$ depends on the total mass density slope, $\gamma$, the luminous matter density slope, $\delta$, and the stellar orbital anisotropy, $\beta$.

In deriving \eqref{eq:cleanmess} several physical assumptions have been made, viz.:
\begin{enumerate}
    \item The creation, destruction and collisions of stars are neglected.
    \item The thin lens approximation applies as the distances from observer to lens and lens to source are far larger than the width of the lens object itself. A projected mass density can then be defined in the lens plane that is responsible for the deflection of light. 
    \item The system is stationary, i.e., $\partial_t =0$. The properties of the lens galaxy vary insignificantly over the short periods in which lensed images are observed. 
    \item The weak field limit of general relativity applies (linearized gravity) with an asymptotically flat (Minkowski) background. For the vast majority of lensing scenarios one assumes $\Phi/c^2 \ll 1$.
    \item Angles of deflection due to gravitational lensing are small, so that the small angles approximation is invoked. 
    \item The Born approximation applies as the deflection angle is small. The gravitational potential along the deflected and undeflected light paths can be considered to be approximately the same.
\end{enumerate}

\subsubsection{Extended Power-Law Model}
For the EPL model \citep{koopmansGravitationalLensingStellar2006}, the most nuanced and complex model considered in this paper, the function $F$ in \eqref{eq:cleanmess} takes the form
\begin{multline}
    F(\gamma,\delta,\beta)= \frac{3-\delta}{(\xi -2\beta)(3-\xi)} \, \times \\  \left[ \frac{\Gamma\qty(\frac{\xi-1}{2})}{\Gamma\qty(\frac{\xi}{2})} - \beta \frac{\Gamma\qty(\frac{\xi +1}{2})}{\Gamma\qty(\frac{\xi +2}{2})} \right] \left[\frac{\Gamma\qty(\frac{\gamma}{2})\Gamma\qty(\frac{\delta}{2})}{\Gamma\qty(\frac{\gamma-1}{2}) \Gamma\qty(\frac{\delta-1}{2})} \right] \, ,
\end{multline}
where $\xi = \gamma + \delta -2$.

Through a specific choice of $\gamma$, $\delta$ and $\beta$, which parameterise the total mass density profile $\rho$, the luminous mass density profile $\nu$ and the stellar orbital anisotropy $\beta$,
\begin{align}
    &\rho (r) = \rho\Z0 (r/r\Z0)^{-\gamma} \, , \\
    &\nu (r) = \nu\Z0 (r/r\Z0)^{-\delta} \, , \\
    &\beta\ = 1 - \sigma^2_\theta / \sigma^2_r \, ,
\end{align} 
simpler models are recovered. Both $\gamma = \gamma_0$ and 
\begin{equation}\label{eq:gammaPar}
    \gamma = \gamma_0 + \gamma_z z_l +\gamma_s \log \Tilde{\Sigma}
\end{equation}
parameterisations are considered for the total matter density profile where the normalised surface mass density $\Tilde{\Sigma} \propto \sigma_0^2 / R_{\rm eff}$. 

\subsubsection{Spherical Power-Law Model}
If both $\delta = 2$ and $\beta = 0$ are fixed, the SPL model is obtained. In this spherically symmetric model, the only free parameter is the total mass density (dark matter halo) profile, $\gamma$, and
\begin{equation}
    F:=F(\gamma,2,0) = \frac{1}{\sqrt{\pi}} \frac{1}{\gamma(3-\gamma)} \, .
\end{equation}

The SPL model is often generalized further to allow for the variation of $\gamma$ with redshift or surface brightness density of the lens galaxy. For our investigation, we limit $\gamma$ to a constant value, $\gamma_0$. However, \citet{chenAssessingEffectLens2019} explore different parametrizations of $\gamma$, which are discussed in Section \ref{sec:Discussion of Results}. 

\subsubsection{Singular Isothermal Sphere Model}
The SIS model is the simplest discussed, where $\gamma = \delta = 2$, $\beta = 0$, and
\begin{equation}
    F:=F(2,2,0) = \frac{1}{2\sqrt{\pi}} \,.
\end{equation}
It assumes that the halo mass of the lens galaxy is in isothermal equilibrium, with gravitational attraction balanced entirely by the pressure associated with the constituent stars' interactions. The total mass density relation is then
\begin{equation}\label{eq:SIS_density}
    \rho (r) =\frac{\sigma^2}{2 \pi G r^2} \,,
\end{equation}
which leads to the distance ratio,
\begin{equation}\label{eq:SIS_d}
    \frac{d_{ls}}{d_s} = \frac{c^2 \theta_E}{4 \pi \sigma_0^2} \, .
\end{equation}

\section{Parameter Determination and Model Preference Results}
\label{sec:Results}
From \eqref{eq:FLRW_d} and \eqref{eq:FLRW_dratio}, or \eqref{eq:Timescape_ratio}, with \eqref{eq:cleanmess} we find a value of $\sigma_0$ that depends purely on model parameters, redshifts $z_s$ and $z_l$, and the Einstein radius $\theta_E$. We can then compare these values with the observationally determined ones via an MCMC sampling procedure, to constrain the defining parameters of both lens and cosmological models. 

The likelihood of the combined lens and cosmological models is determined via 
\begin{equation}
    \chi^2 = \sum^{161}_{i=1} \frac{\left( \sigma_0 - \sigma_0^{\text{model}}(z_l, z_s, \theta_E, \mathbf{p}) \right)^2}{\Delta \sigma_0^2} \, ,
\end{equation}
where $\mathbf{p}$ are the defining parameters for a specific lens model i.e., $\gamma$, $\delta$ and $\beta$ as well as cosmological parameters $\Omega_{\rm M0}$ and $f_{\rm v0}$. $\sigma_0^{\text{model}}$ is given by the combination of lens and cosmological models.

We initially take wide uniform priors for all parameters to avoid biases in the values determined.The stellar orbital anisotropy $\beta$ of all lens galaxies in the catalogue is not measured, so we assume a $2\sigma$ Gaussian prior $\beta = 0.18 \pm 0.26$, established from a study of nearby elliptical galaxies and adopted by \citet{chenAssessingEffectLens2019}. A full list of the priors involved in our MCMC sampling is given in \ref{tab:MCMC_priors}.

\begin{figure*} 
     \centering
     \begin{subfigure}[b]{0.6\textwidth}
         \centering
         \includegraphics[width=\textwidth]{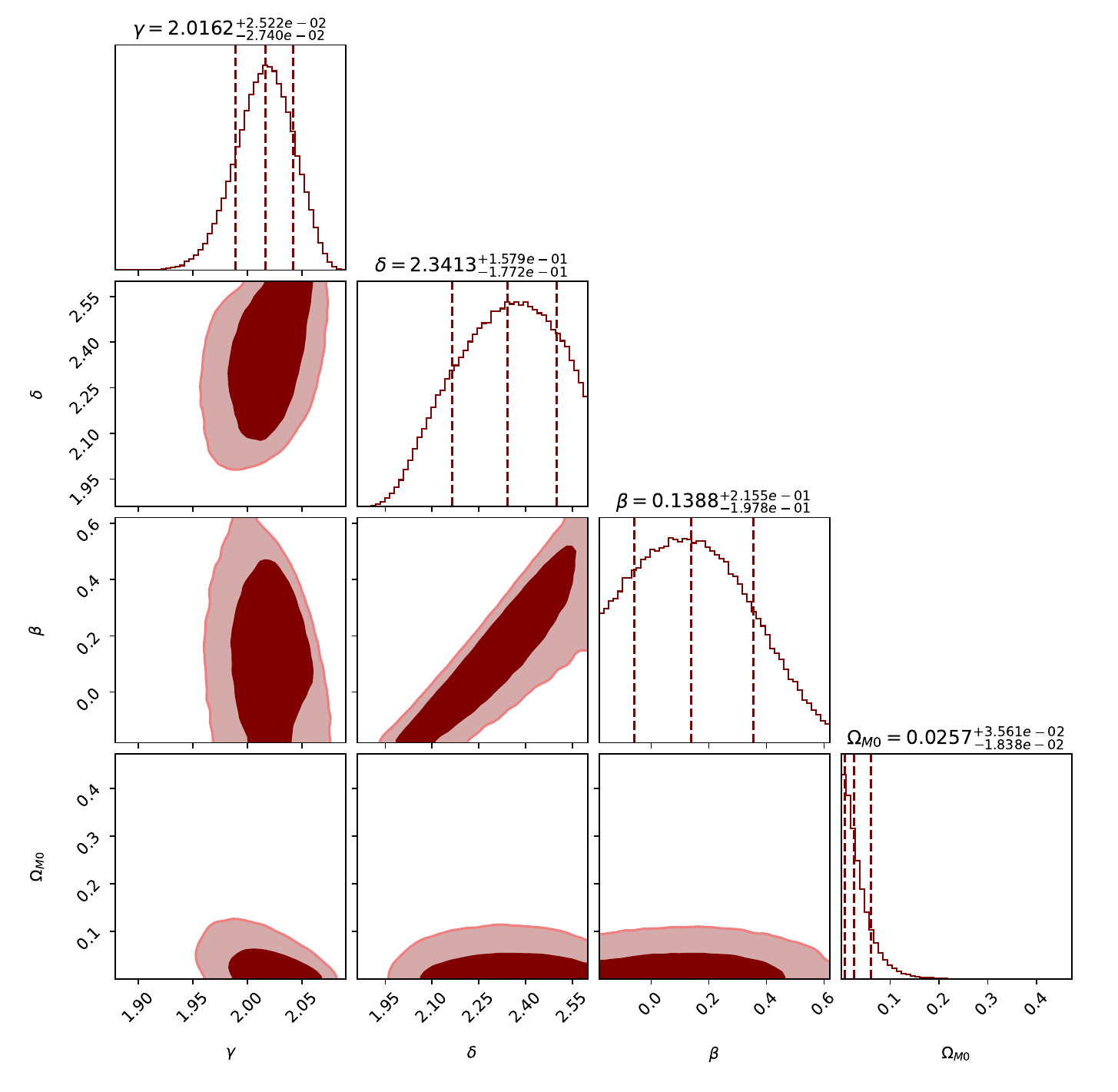}
         \caption{$k=0$ FLRW}
     \end{subfigure}
     \hfill
     \begin{subfigure}[b]{0.6\textwidth}
         \centering
         \includegraphics[width=\textwidth]{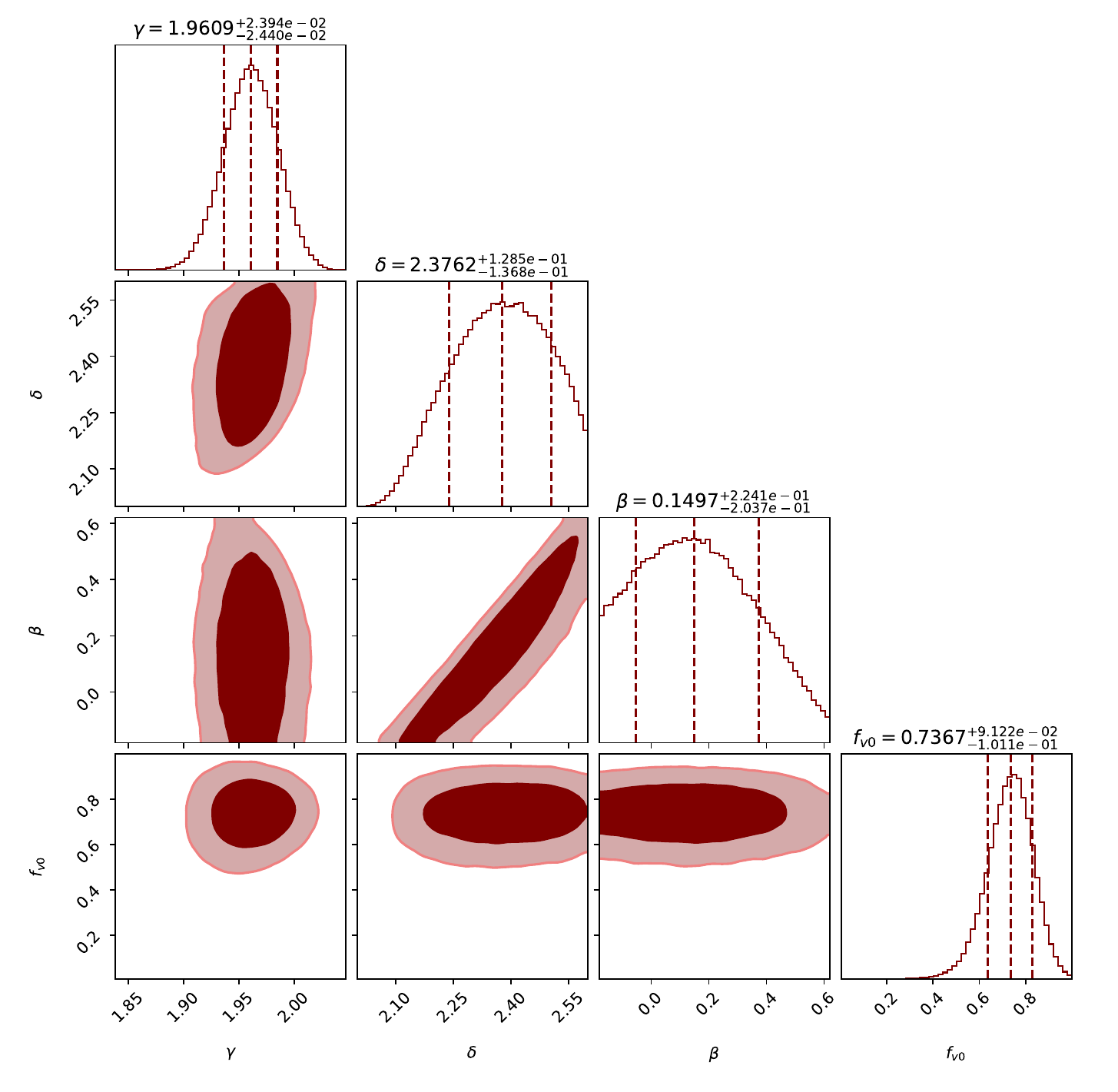}
         \caption{Timescape}
     \end{subfigure}
     \caption{Parameter probability distributions of the EPL lens model for both spatially flat FLRW and timescape models, constrained using 161 lensing systems. Dashed lines representing the $2\sigma$ bounds and median value of MCMC samples. }
     \label{fig:EPL MCMC}
\end{figure*}

\begin{figure*} 
     \centering
     \begin{subfigure}[b]{0.96\textwidth}
         \centering
         \includegraphics[width=\textwidth]{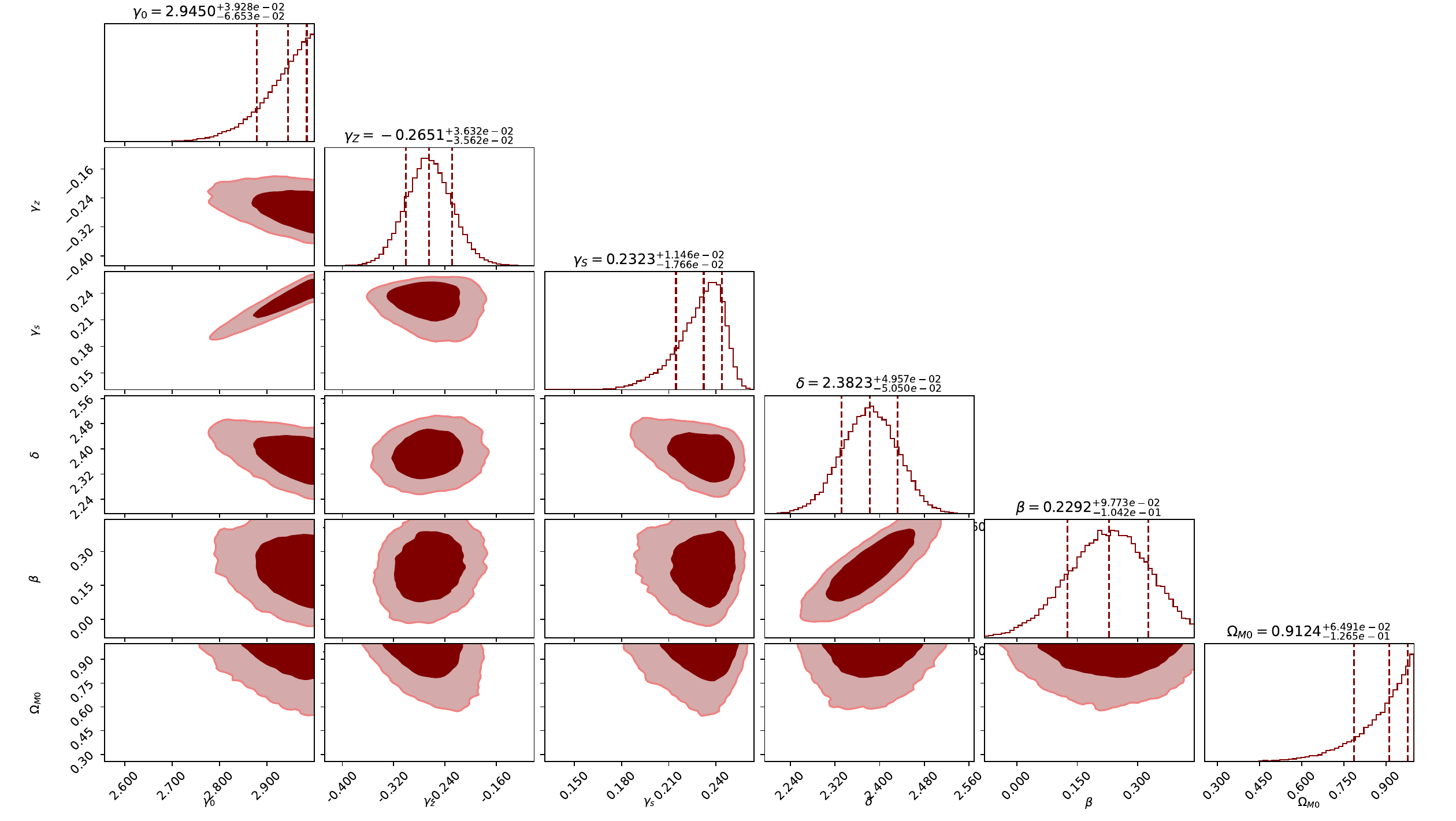}
         \caption{$k=0$ FLRW}
     \end{subfigure}
     \hfill
     \begin{subfigure}[b]{0.96\textwidth}
         \centering
         \includegraphics[width=\textwidth]{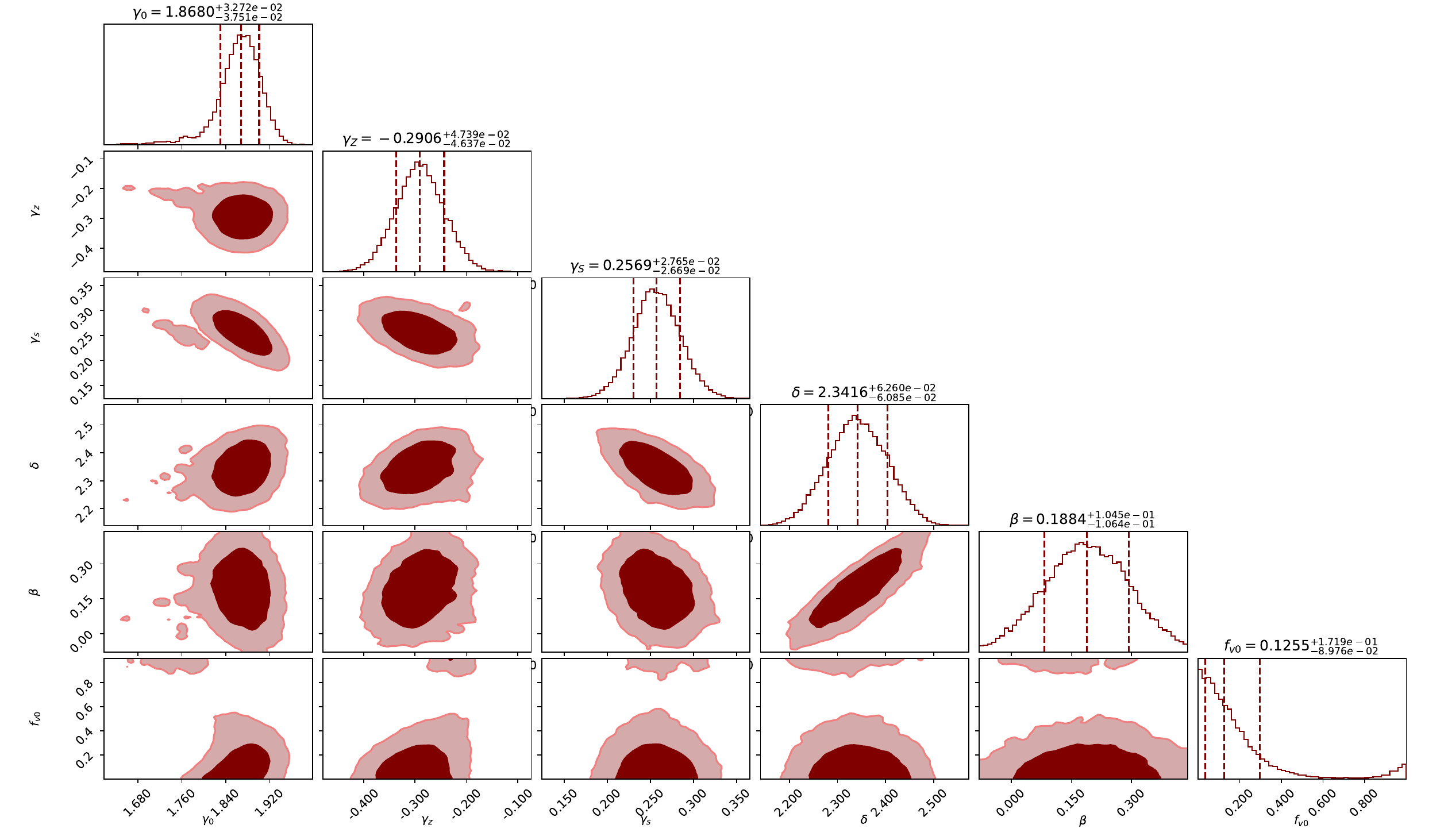}
         \caption{Timescape}
     \end{subfigure}
     \caption{Parameter probability distributions of the fully extended EPL lens model for both spatially flat FLRW and timescape models, constrained using 161 lensing systems. }
     \label{fig:EPL_full_MCMC}
\end{figure*}

To obtain the results given in figs \ref{fig:EPL MCMC} and \ref{fig:EPL_full_MCMC}, the MCMC sampler ran for 50,000 steps with 32 walkers and a \emph{burn-in} phase discarding the first 20\% of the samples. These were created using the \emph{emcee} python package.   

To distinguish between the spatially flat FLRW and timescape models, which have the same number of free parameters, a straightforward Bayes factor, $B$, can be calculated by integrating the likelihoods $\mathcal{L} \sim e^{-\chi^2/2}$ over the $2\sigma$ priors $f_{\rm v0} \in (0.5,0.799)$, $\Omega_{\rm M0} \in (0.143,0.487)$ \citep{damApparentCosmicAcceleration2017,trottaBayesianMethodsCosmology2017} and corresponding lens model parameters. The cosmological parameter priors are determined using non-parametric fits of the Planck CMB data \citep{aghamousaNonparametricTestConsistency2015} constrained by estimates of the angular BAO scale from the BOSS survey data and Lyman alpha forest statistics \citep{alamClusteringGalaxiesCompleted2017, delubacBaryonAcousticOscillations2015}. Following \citet{damApparentCosmicAcceleration2017} we use wide priors for both models so as not to unfairly disadvantage \LCDM. The priors usually adopted for $\Lambda$CDM, including a precise estimate $\Omega_{\rm M0}=0.315 \pm 0.007$ \citep{planckcollaborationPlanck2018Results2020}, arise from constraints on perturbation theory on an FLRW background. A timescape equivalent has yet to be developed to constrain the equivalent free parameter, $f_{\rm v0}$. We take the luminosity density profile $\delta$ to have priors informed by profiles fitted to Hubble Space Telescope imagery, giving $2.003<\delta<2.343$. The full list of priors used for Bayesian comparison are given in \ref{tab:Bayes_Priors}.

We adopt the Jeffrey's scale for the Bayes factor
\begin{equation}\label{eq:Bayes}
    B = \frac{\int \mathcal{L}^{\rm timescape} \, \dd f_{\rm v0} \, \dd \gamma \, \dd \delta \, \dd \beta}{\int \mathcal{L}^{\rm FLRW} \, \dd \Omega_{\rm M0} \, \dd \gamma \, \dd \delta \, \dd \beta}
\end{equation}
so that values of $B>1$ will indicate preference for timescape and $B<1$ for FLRW. By the standard interpretation, evidence with $|\ln B|<1$ is `not worth more than a bare mention' \citep{kr} or `inconclusive' \citep{t07}, while $1\leq|\ln B|< 3$, $3\leq|\ln B|<5$ and $|\ln B|\ge5$ indicate `positive', `strong' and
`very strong' evidences respectively \citep{kr}.

\begin{table}
\begin{tabular}{ |p{0.28\columnwidth}||p{0.28\columnwidth}|p{0.28\columnwidth}| }
\hline
\multicolumn{3}{|c|}{Bayes Factors (B)} \\
\hline
Lens Model& $B$ &$\abs{\ln{B}}$ \\
\hline
EPL   & 0.4928 & 0.7077 \\
SPL& $9.38 \times 10^{-7}$  & 13.9 \\
SIS & $3.28 \times 10^{-3}$ & 5.72  \\

\end{tabular}
\caption{Bayes factors \eqref{eq:Bayes}, where $B < 1$ favours the spatially flat FLRW model.}
\label{tab:Bayes Factors}
\end{table}

\renewcommand{\arraystretch}{1.5}
\begin{table*}
\begin{tabular}{ |p{3cm}||p{3cm}|p{3cm}|p{3cm}|p{3cm}|  }
 \hline
 \multicolumn{5}{|c|}{FLRW ($k=0$)} \\
 \hline
 Lens Model& $\beta$ &$\delta$ &$\gamma$ &$\Omega_{\rm M0}$\\
 \hline
 EPL (full parametrisation)*   &  $0.2292^{+9.773 \times 10^{-2}}_{-1.042 \times 10^{-1}}$ & $2.3823^{+4.947 \times 10^{-2}}_{-5.050 \times 10^{-2}}$ & $\gamma_0$ = $2.9450^{+3.928 \times 10^{-2}}_{-6.653 \times 10^{-2}}$ & $0.9124^{+6.491\times 10^{-2}}_{-1.265 \times 10^{-1}}$ \\
 & & & $\gamma_z$ = $-0.2651^{+3.632 \times 10^{-2}}_{-3.562 \times 10^{-2}}$  &  \\
  & & & $\gamma_s$ = $0.2323^{+1.146 \times 10^{-2}}_{-1.766 \times 10^{-2}}$  &  \\
 EPL ($\gamma = \gamma_0$)  & $0.1388^{+2.155 \times 10^{-1}}_{-1.978 \times 10^{-1}}$  & $2.3413^{+1.579 \times 10^{-1}}_{-1.722 \times 10^{-1}}$ & $2.0162^{+2.522 \times 10^{-2}}_{-2.740 \times 10^{-2}}$  & 
$0.0257^{+3.561 \times 10^{-2}}_{-1.838 \times 10^{-2}}$ \\
 SPL&  --  & --   & $1.9306^{+1.707 \times 10^{-2}}_{-1.844 \times 10^{-2}}$ & $0.0188^{+2.317 \times 10^{-2}}_{-1.319 \times 10^{-2}}$ \\
 SIS & -- & -- &  -- & $0.0036^{+5.00 \times 10^{-3}}_{-2.70 \times 10^{-3}}$  \\
\end{tabular}
\caption{Ideal parameter values obtained from MCMC sampling of velocity dispersions obtained using FLRW distances and one of the listed lens models. }
\label{tab:FLRW MCMC}
\end{table*}

\begin{table*}
\begin{tabular}{ |p{3cm}||p{3cm}|p{3cm}|p{3cm}|p{3cm}|  }
\hline
\multicolumn{5}{|c|}{Timescape} \\
\hline
Lens Model& $\beta$ &$\delta$ &$\gamma$ &$f_{\rm v0}$\\
\hline
 EPL (full parametrisation)*   &  $0.2292^{+9.773 \times 10^{-2}}_{-1.042 \times 10^{-1}}$ & $2.3823^{+4.947 \times 10^{-2}}_{-5.050 \times 10^{-2}}$ & $\gamma_0$ = $2.9450^{+3.928 \times 10^{-2}}_{-6.653 \times 10^{-2}}$ & $0.9124^{+6.491\times 10^{-2}}_{-1.265 \times 10^{-1}}$ \\
 & & & $\gamma_z$ = $-0.2651^{+3.632 \times 10^{-2}}_{-3.562 \times 10^{-2}}$  &  \\
  & & & $\gamma_s$ = $0.2323^{+1.146 \times 10^{-2}}_{-1.766 \times 10^{-2}}$  &  \\
EPL ($\gamma = \gamma_0$)  & $0.1497^{+2.241 \times 10^{-1}}_{-2.037 \times 10^{-1}}$ & $2.3762^{+1.285 \times 10^{-1}}_{-1.368 \times 10^{-1}}$ &   $1.9609^{+2.394 \times 10^{-2}}_{-2.740 \times 10^{-2}}$ & $0.7367^{+9.122 \times 10^{-2}}_{-1.011 \times 10^{-1}}$ \\
SPL&  --  & -- &$1.8837^{+6.20 \times 10^{-3}}_{-6.20 \times 10^{-3}}$ & 
 $0.6751^{+1.73 \times 10^{-2}}_{-1.73 \times 10^{-2}}$ \\
SIS & -- & -- &  -- &  $0.7044^{+3.780 \times 10^{-2}}_{-3.859 \times 10^{-2}}$   \\
\end{tabular}
\caption{Ideal parameter values obtained from MCMC sampling of velocity dispersions obtained using timescape distances and one of the listed lens models.  }
\label{tab:Timescape MCMC}
\end{table*}

\subsection{Discussion of Results}\label{sec:Discussion of Results}
The Bayes factors favour a spatially flat FLRW model over the timescape model in all cases but with varying degrees of strength, as shown in Table~\ref{tab:Bayes Factors}. For the SIS and SPL lens models, the preference is a strong. However, both FLRW and timescape have a minimum $\chi^2$ per degree of freedom $\sim2$, which shows in both cases the fit could be improved, specifically within the choice and parametrisation of the lens model.  It is important to note that the Bayes factors of Table~\ref{tab:Bayes Factors} should not be interpreted na\"{\i}vely, as lower $\chi^2$ for FLRW models comes at the expense of an unphysical matter density at the extremes $\Omega_{\rm M0}\simeq0$ or $\Omega_{\rm M0}\simeq1$. By contrast, the values of $f_{\rm v0}$ predicted are within the $2\sigma$ priors $f_{\rm v0} \in (0.5,0.799)$ for timescape and remain physically plausible for the majority of lens model parametrisations. 

\citet{chenAssessingEffectLens2019} already noted that $\Omega_{\rm M0}\simeq 0$ for specific lens models. However, by considering
\begin{enumerate}[topsep=0pt, itemsep=0pt,leftmargin=*, align=right] 
    \item an alternate parametrisation of $\gamma$ to include a dependency on redshift and normalised surface mass density of each lens galaxy, $\gamma = \gamma_0 + \gamma_z  z_l + \gamma_s  \log{\Tilde{\Sigma}}$; and
    \item $\delta$ as an observable for each lens galaxy
\end{enumerate} 
more realistic values of $\Omega_{\rm M0}$ are inferred. One can only fit $\delta$ given high-resolution imagery of a lens galaxy; for the \citet{chenAssessingEffectLens2019} catalogue this reduces the sample of lensing systems from 161 to 130. Future surveys are predicted to observe several orders of magnitude more lensing systems. The requirement of follow up high resolution imaging will therefore face significantly greater challenges on account of the vast increase of data volume. When fitting a global value for $\delta$ in the EPL model, we find values consistent with $2\sigma$ priors informed by HST imagery of $2.003 < \delta < 2.343$ as shown in Fig \ref{fig:EPL MCMC} and \ref{fig:EPL_full_MCMC}.

Physically plausible values of $\Omega_{\rm M0}$ for FLRW have only been found when steps (i) and (ii) are applied to the further parametrised EPL model. If $\delta$ is not constrained for each system in the EPL model, then $\Omega_{\rm M0}\simeq1$, the other unphysical extreme. Timescape also returns the physically implausible extreme value for $f_{\rm v0} \simeq 0$ which corresponds exactly to the Einstein de~Sitter universe ($\Omega_{\rm M0} = 1$) for the full paramaterisation of $\gamma$. 

\section{Mock Catalogues and Parameter Fitting}
\label{sec:Simulations}
\subsection{Methodology}
The unphysical cosmological parameter values found through MCMC sampling, in particular of $\Omega_{\rm M0}$, motivate further investigation. Thus, we generate mock catalogues to gauge how sensitive the fitting procedure is on cosmological parameters. We generate mock catalogues using the following procedure:
\begin{enumerate}
    \item Using the catalogue data for the 161 lensing systems, we apply the relation for the simplest case lens model SIS \eqref{eq:SIS_d}, in combination with distance ratios found from cosmology (see  \eqref{eq:FLRW_d} and \eqref{eq:FLRW_dratio} or \eqref{eq:Timescape_ratio}) to find the \emph{model ideal} values of $\sigma_0$. This requires an initial seed value of $\Omega_{\rm M0}$ or $f_{\rm v0}$, which we hope to later recover when fitting. 
    \item Gaussian noise is then added to $\sigma_0$ for each of the 161 lens galaxies. 
    \item We then use \eqref{eq:SIS_d} with the catalogue data and the velocity dispersion, with added noise, to find a new value of the distance ratio $d_{ls}/d_s$.
    \item The mock distance ratio and the value determined from either the timescape or spatially-flat FLRW models can then be compared through the $\chi^2$ test 
    \begin{equation}
    \chi^2 = \sum^{161}_{i=1} \qty( \frac{\mathcal{D}_i^{\rm model}(\mathbf{q}) - \mathcal{D}_i^{\rm mock}(\mathbf{q}^{\text{seed}}) }{\Delta \mathcal{D}_i})^2 \,,
    \end{equation}
    where $\mathbf{q} = \{\Omega_{\rm M0}$, $f_{\rm v0}$\} depending on cosmology.
    \item The $\Omega_{\rm M0}$ and $f_{\rm v0}$ parameters are varied, the best fit values corresponding to the minimized $\chi^2$ value. 
    \item We repeat this for $10^4$ mock samples, with the best fit parameters for each cosmology being binned into a histogram for each individual mock. We then determine whether the original seed parameters of $\Omega_{\rm M0}$ and $f_{\rm v0}$ are recovered from the fitting procedure. 
\end{enumerate}
A wide variety of seed values are chosen with parameters in the range $\Omega_{\rm M0}\in\{0.1,0.9\}$ for spatially flat FLRW, and similarly $f_{\rm v0}\in\{0.1,0.9\}$ for timescape. We find that regardless of the initial input values of the cosmological parameters, this procedure always lowers $\Omega_{\rm M0}$ significantly for FLRW. For timescape, a value of $f_{\rm v0} \simeq 0.73$ is returned irrespective of the input value.

\subsection{Simulation Results and Discussion}

\begin{figure}
    \includegraphics[width=\columnwidth]{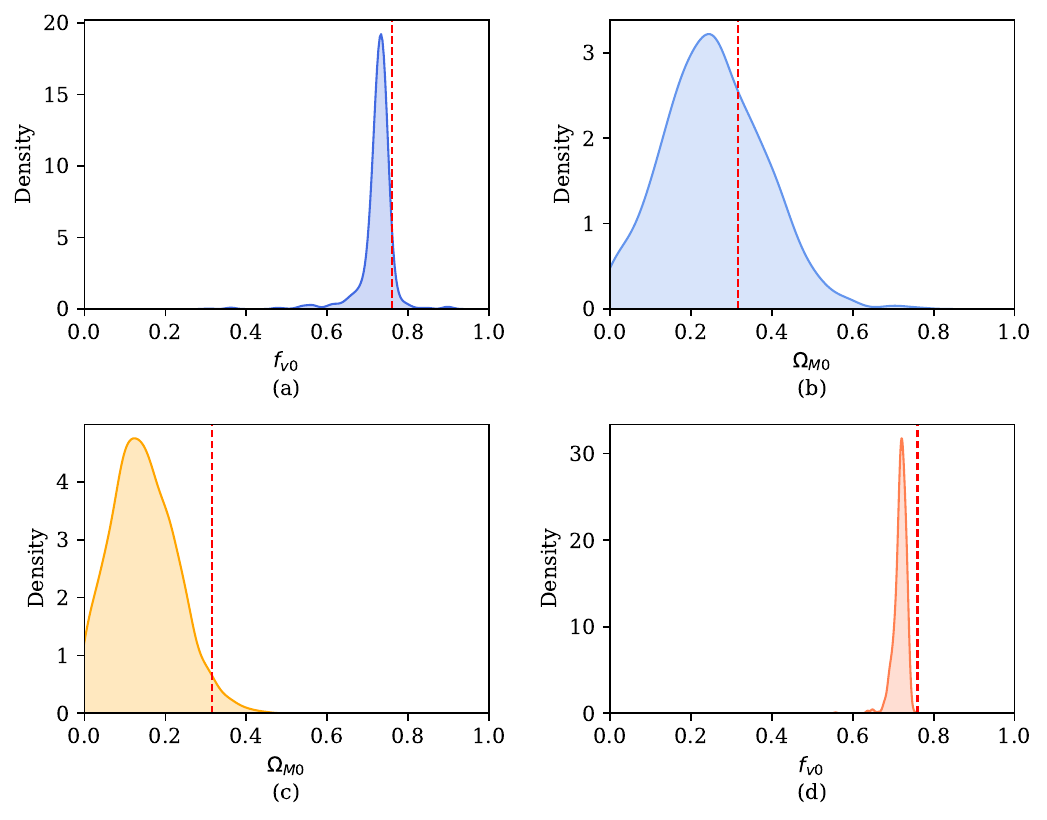}
    \caption{Kernel density estimate plots of the best fit parameters for: (a) timescape fit to timescape data, (b) FLRW fit to timescape data, (c) FLRW fit to FLRW data and (d) timescape fit to FLRW data. The dashed vertical red line indicates the initial parameter value used to generate the mock data, $f_{\rm v0}=0.76$ and $\Omega_{\rm M0}=0.315$ for timescape and FLRW models respectively.}
    \label{fig:KDE fits}
\end{figure}

\begin{figure*}
     \centering
     \begin{subfigure}[b]{\columnwidth}
         \centering
         \includegraphics[width=\textwidth]{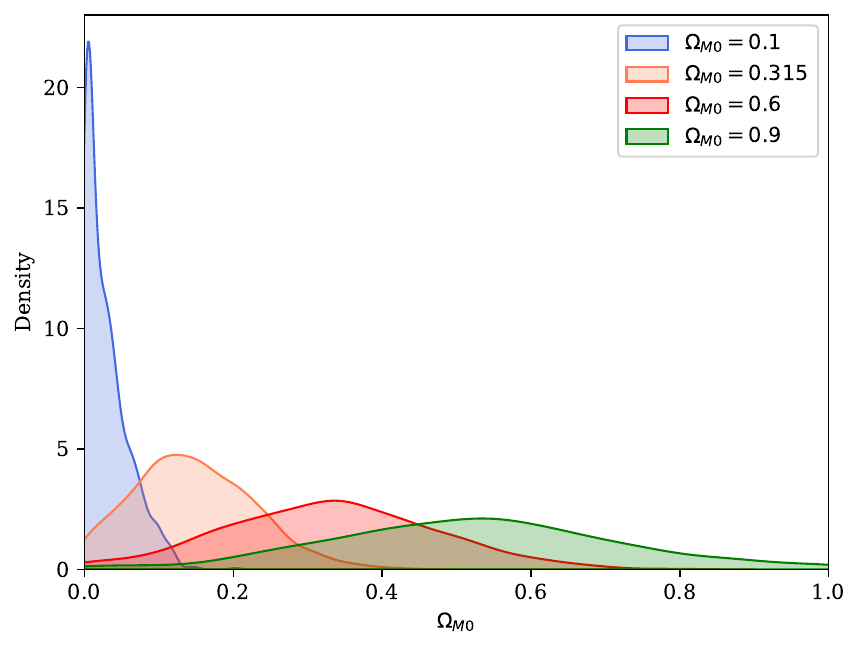}
         \caption{\textbf{FLRW}}
         \label{fig:FLRW KDE comp}
     \end{subfigure}
     \hfill
     \begin{subfigure}[b]{\columnwidth}
         \centering
         \includegraphics[width=\textwidth]{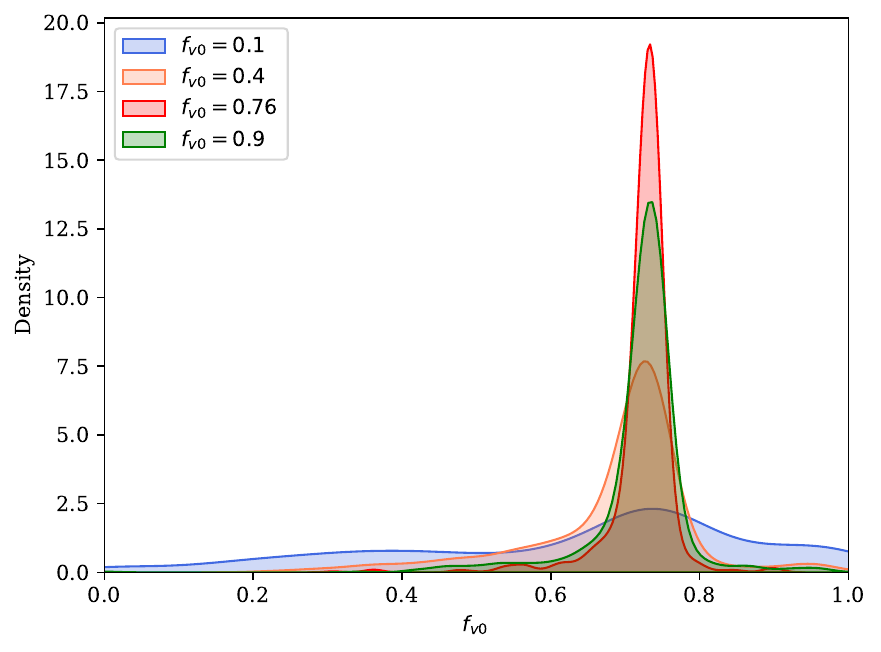}
         \caption{\textbf{Timescape}}
         \label{fig:Time KDE comp}
     \end{subfigure}
     \caption{Kernel density estimate plots showing the best fit cosmological parameters for both FLRW (a) and timescape (b) models fit against data generated with varying initial parameter values. }
     \label{fig:KDE comp}
\end{figure*}

The histograms produced from the mock data provide interesting results, shown in Fig~\ref{fig:KDE fits}. Na\"{\i}vely, one would assume that if the cosmological model plays a significant role in the pipeline then the generated cosmological parameters would match the input values.  However, this is only the case when the uncertainty in the velocity dispersion is set artificially low.

The timescape fit finds values of $f_{\rm v0} \approx 0.73$ which are consistent with the model expectation based on constraints from SneIa distances, CMB, etc~\citep{DuleyTimescapeRadiation2013}. In fact, this is still true for timescape fits of the FLRW mocks. The FLRW mocks have a significantly wider cosmological parameter spread, as shown in Fig.~\ref{fig:FLRW KDE comp} as compared to Fig.~\ref{fig:Time KDE comp}. Furthermore, the FLRW model yields an $\Omega_{\rm M0}$ more in-line with the expected value of $\Omega_{\rm M0}$ from the Planck CMB data \citep{planckcollaborationPlanck2018Results2020} when it is fit against data generated with the timescape model. 
For FLRW the maximum likelihood of $\Omega_{\rm M0}$ is found to increase as the input $\Omega_{\rm M0}$ is lowered, resulting in an overall maximum likelihood $\Omega_{\rm M0}\to 0$, corresponding to an unphysical Milne universe. For timescape the maximum likelihood increases for seed values close to $f_{\rm v0} \simeq 0.73$. 

\begin{figure}
    \centering
    \includegraphics[width=0.8\columnwidth]{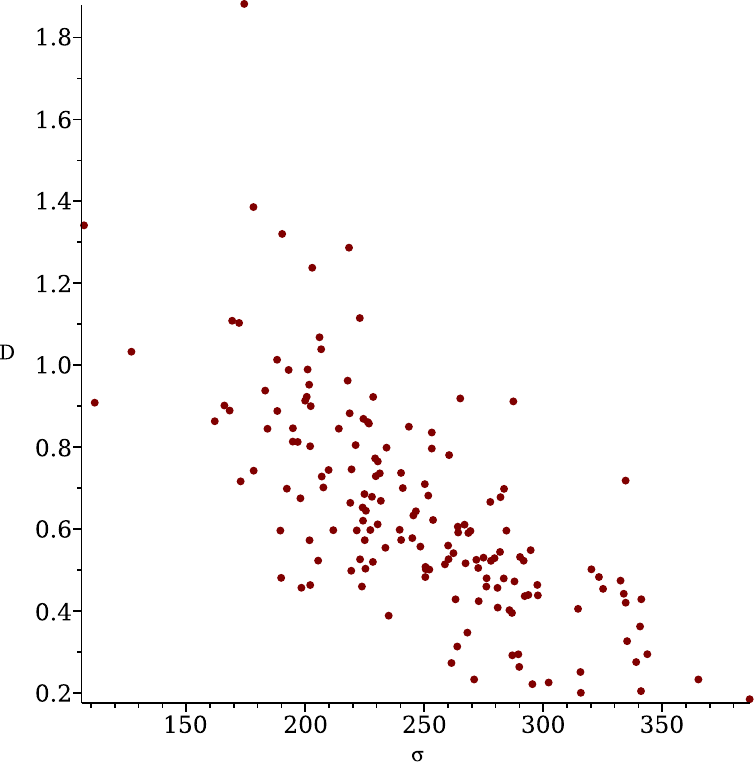}
    \caption{Distance ratios against velocity dispersion ($\rm kms^{-1}$) for a Single Isothermal Sphere lens model using catalogue data for 161 systems.}
    \label{fig:Dratio_veldis}
\end{figure}

Both the data and mock catalogues produce some outlying unphysical distance ratios $\mathcal{D} > 1$ with closer source than lens to observer (see Fig.~\ref{fig:Dratio_veldis}).
This is an artefact of choosing a global power-law lens model and fitting to a data set with high variations in the observed velocity dispersions. Some systems are not well described by the lens model choice, and thus produce abnormally high distance ratios that skew the distribution. Fitting any curve $D(\sigma)$ to the data of Fig.~\ref{fig:Dratio_veldis} will be skewed by the unphysical values.

\begin{figure*}
     \centering
     \begin{subfigure}[b]{\columnwidth}
         \centering
         \includegraphics[width=0.95\textwidth]{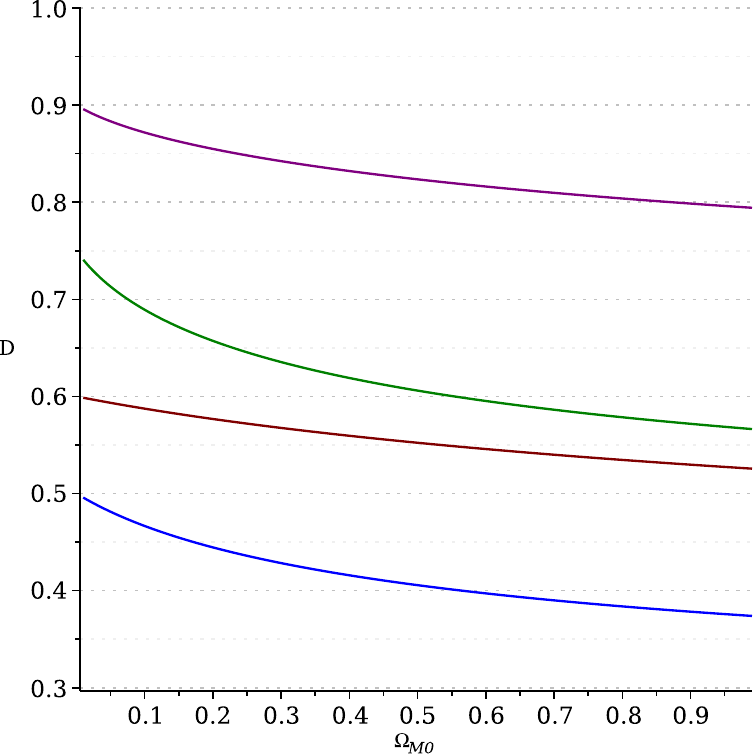}
        \caption{\textbf{FLRW}}
         \label{fig:Dratio_FLRW}
     \end{subfigure}
     \begin{subfigure}[b]{\columnwidth}
         \centering
         \includegraphics[width=0.95\textwidth]{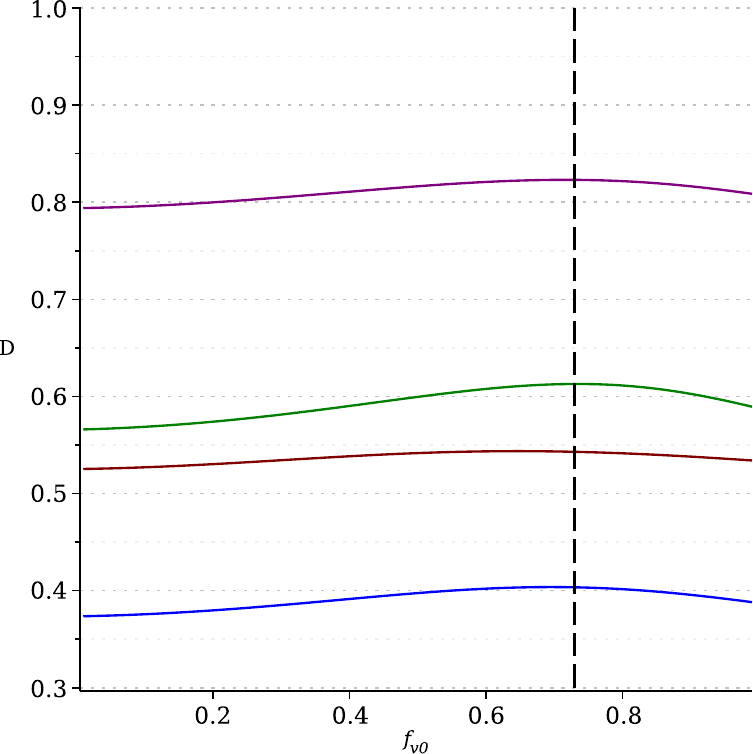}
         \caption{\textbf{Timescape}}
         \label{fig:Dratio_time}
     \end{subfigure}
     \caption{Distance ratios for fixed lens and source redshifts in both spatially flat FLRW and timescape models with varying $\Omega_{M0}$ and $f_{v0}$ respectively.}
     \label{fig:MapleDratioFig}
\end{figure*}

The implications can by understood by considering Fig.~\ref{fig:Dratio_FLRW} and Fig.~\ref{fig:Dratio_time} where distance ratios $\mathcal{D}^{\text{FLRW}}$ and $\mathcal{D}^{\text{Time}}$ are shown for particular fixed source and lens redshifts, $z_s$ and $z_l$. Varying $z_s$ and $z_l$ we see that while the distance ratio generally increases with increasing $z_s$, the maximum is always found at $\Omega_{\rm M0} \simeq 0$ for FLRW and $f_{\rm v0} \simeq 0.73$ for timescape.
With a high enough uncertainty in the velocity dispersion, much larger distance ratios are possible in the simulations. Therefore, the fitting of cosmological parameters is forced towards the values that enable the highest possible distance ratios. 

Whilst our simulations were performed using a SIS lens model, the entire class of lens models in our investigation return unphysical distance ratios, $\mathcal{D} \geq 1$, when using the ideal parameters determined by MCMC sampling. Even the EPL model with the full parametrisation \eqref{eq:gammaPar}, which enables a slight variation of the power law model between systems, often yields unphysical distance ratios. 

       \section{Conclusions}
\label{sec:Conclusions}
In order to increase the goodness of fit of power-law lens models, additional empirical parameters are often added to the lens models along with constraints from high resolution imagery. The simplest parametrisations of these models, in combination with the standard spatially flat FLRW cosmology, have already been shown to produce poor fits to data without additional observationally determined luminosity density profiles \citep{chenAssessingEffectLens2019}. We find that all choices of power-law model investigated na\"{\i}vely prefer a spatially flat FLRW cosmology, but as a consequence delegate all the mass along the line of sight to the lens galaxy, resulting in $\Omega_{\rm M0} \simeq 0$. For the full parametrisation \eqref{eq:gammaPar} of the EPL model the fits return an Einstein de~Sitter universe with $\Omega_{\rm M0} = 1$,  which is also clearly unphysical.

The timescape cosmology gives void fraction values of $0.68 \leq f_{\rm v0} \leq 0.74$ consistent with constraints from independent tests \citep{DuleyTimescapeRadiation2013,LaneCosmologicalFoundations2023} in all cases except that of \eqref{eq:gammaPar} when it also returns the same unphysical Einstein de~Sitter universe ($f_{\rm v0}=0$).
Simulations with a fixed SIS lens model consistently return values of $f_{\rm v0} \simeq 0.73$, regardless of the input seed value. Our results highlight the imperative of exploring alternatives to the standard \LCDM cosmology. Even if any particular non-FLRW cosmology is incorrect, such comparisons may give insight into breaking the many degeneracies present in SGL analyses to date. 

It is clear that either 
\begin{enumerate}[topsep=0pt, itemsep=0pt,leftmargin=*, align=right] 
    \item power-law models cannot be applied globally to large surveys of lensing systems as they make unphysical predictions of the distance ratios, or,
    \item our measurements of the velocity dispersions of lens galaxies are not currently at a high enough accuracy and precision to enable this kind of analysis.
\end{enumerate}

Navarro-Frenk-White profiles are generally taken as the standard for modelling the dark matter haloes of elliptical galaxies, rather than the power-law models used here for large SGL catalogues. The use of a general lens model across all systems in the catalogue, assuming that on average they will be well-defined, is evidently not the case. Many lens galaxies have large uncertainties in their respective velocity dispersion measurements, leading to unphysical distance ratios for many lensing systems regardless of the choice of power-law used.

Systems that deviate from these power-law models negatively skew the distribution of distance ratios, leading to biasing in the determination of cosmological parameters. In fact, even an individualised EPL model, with the full parametrisation of $\gamma$, fails to produce physically plausible distance ratios for all systems. What results is that the `best fit' cosmological parameters found are those which can produce the highest possible distance ratios to match the unphysical predictions of the lens models. 

Without additional imaging data, the technique of using distance ratios as a test statistic is weak at constraining cosmological parameters. However, it has the advantage of a far larger dataset than that of using time-delay distances. With the next generation of telescopes (JWST, Euclid, LSST etc) predicted to observe even more lensing systems, it is of paramount importance that appropriate lens models are determined in advance of the upcoming catalogues if one aims to use strong lensing for robust statistical constraints in cosmology.

Time-delay cosmography is able to constrain the lens mass distribution with far greater precision than the distance sum rule test. This arises from the extra constraints given by the magnification and time-delay between each of the images. In addition, time-delay cosmography models lensing systems individually, whereas the distance ratio technique fits global lens models to large catalogues of highly variable systems. At present, despite the smaller number of systems available, time-delay cosmography may present the only viable way to use strong gravitational lensing to constrain both cosmological models and lens matter models with the precision of measurements available in ongoing surveys.

\section*{Acknowledgements}

We warmly thank Jenny Wagner for her perceptive physical insights in defining this project, and for introducing us to the concepts, methodology, statistical and systematic issues that dominate cosmological applications of strong gravitational lensing. We also thank Zachary Lane, Marco Galoppo, Morag Hills and Michael Williams for our many discussions and their helpful comments.
\section*{Data Availability}

The data used for this analysis was taken from the catalogue of lensing surveys compiled in \citet{chenAssessingEffectLens2019}.
\newpage

\bibliographystyle{mnras}
\bibliography{main}

\begin{thebibliography}{}
\makeatletter
\relax
\def\mn@urlcharsother{\let\do\@makeother \do\$\do\&\do\#\do\^\do\_\do\%\do\~}
\def\mn@doi{\begingroup\mn@urlcharsother \@ifnextchar [ {\mn@doi@}
  {\mn@doi@[]}}
\def\mn@doi@[#1]#2{\def\@tempa{#1}\ifx\@tempa\@empty \href
  {http://dx.doi.org/#2} {doi:#2}\else \href {http://dx.doi.org/#2} {#1}\fi
  \endgroup}
\def\mn@eprint#1#2{\mn@eprint@#1:#2::\@nil}
\def\mn@eprint@arXiv#1{\href {http://arxiv.org/abs/#1} {{\tt arXiv:#1}}}
\def\mn@eprint@dblp#1{\href {http://dblp.uni-trier.de/rec/bibtex/#1.xml}
  {dblp:#1}}
\def\mn@eprint@#1:#2:#3:#4\@nil{\def\@tempa {#1}\def\@tempb {#2}\def\@tempc
  {#3}\ifx \@tempc \@empty \let \@tempc \@tempb \let \@tempb \@tempa \fi \ifx
  \@tempb \@empty \def\@tempb {arXiv}\fi \@ifundefined
  {mn@eprint@\@tempb}{\@tempb:\@tempc}{\expandafter \expandafter \csname
  mn@eprint@\@tempb\endcsname \expandafter{\@tempc}}}

\bibitem[\protect\citeauthoryear{Aghamousa \& Shafieloo}{Aghamousa \&
  Shafieloo}{2015}]{aghamousaNonparametricTestConsistency2015}
Aghamousa A.,  Shafieloo A.,  2015, \mn@doi [Journal of Cosmology and
  Astroparticle Physics] {10.1088/1475-7516/2015/06/003}, 06, 003

\bibitem[\protect\citeauthoryear{Aghanim et~al.,}{Aghanim
  et~al.}{2020}]{planckcollaborationPlanck2018Results2020}
Aghanim N.,  et~al., 2020, \mn@doi [Astronomy \& Astrophysics]
  {10.1051/0004-6361/201833910}, 641, A6

\bibitem[\protect\citeauthoryear{Alam et~al.,}{Alam
  et~al.}{2017}]{alamClusteringGalaxiesCompleted2017}
Alam S.,  et~al., 2017, \mn@doi [Monthly Notices of the Royal Astronomical
  Society] {10.1093/mnras/stx721}, 470, 2617

\bibitem[\protect\citeauthoryear{Aluri et~al.,}{Aluri
  et~al.}{2023}]{aluriObservableUniverseConsistent2023}
Aluri P.~K.,  et~al., 2023, \mn@doi [Classical and Quantum Gravity]
  {10.1088/1361-6382/acbefc}, 40, 094001

\bibitem[\protect\citeauthoryear{Auger, Treu, Bolton, Gavazzi, Koopmans,
  Marshall, Bundy  \& Moustakas}{Auger et~al.}{2009}]{augerSloanLensACS2009}
Auger M.~W.,  Treu T.,  Bolton A.~S.,  Gavazzi R.,  Koopmans L. V.~E.,
  Marshall P.~J.,  Bundy K.,   Moustakas L.~A.,  2009, \mn@doi [Astrophysical
  Journal] {10.1088/0004-637X/705/2/1099}, 705, 1099

\bibitem[\protect\citeauthoryear{Auger, Treu, Bolton, Gavazzi, Koopmans,
  Marshall, Moustakas  \& Burles}{Auger et~al.}{2010}]{augerSloanLensACS2010}
Auger M.~W.,  Treu T.,  Bolton A.~S.,  Gavazzi R.,  Koopmans L. V.~E.,
  Marshall P.~J.,  Moustakas L.~A.,   Burles S.,  2010, \mn@doi [Astrophysical
  Journal] {10.1088/0004-637X/724/1/511}, 724, 511

\bibitem[\protect\citeauthoryear{Bernardi et~al.,}{Bernardi
  et~al.}{2003}]{bernardiEarlyTypeGalaxiesSloan2003}
Bernardi M.,  et~al., 2003, \mn@doi [Astronomical Journal] {10.1086/367776},
  125, 1817

\bibitem[\protect\citeauthoryear{Bolton, Burles, Koopmans, Treu, Gavazzi,
  Moustakas, Wayth  \& Schlegel}{Bolton et~al.}{2008}]{boltonSloanLensACS2008}
Bolton A.~S.,  Burles S.,  Koopmans L. V.~E.,  Treu T.,  Gavazzi R.,  Moustakas
  L.~A.,  Wayth R.,   Schlegel D.~J.,  2008, \mn@doi [Astrophysical Journal]
  {10.1086/589327}, 682, 964

\bibitem[\protect\citeauthoryear{Brownstein et~al.,}{Brownstein
  et~al.}{2012}]{brownsteinBOSSEmissionLineLens2012}
Brownstein J.~R.,  et~al., 2012, \mn@doi [Astrophysical Journal]
  {10.1088/0004-637X/744/1/41}, 744, 41

\bibitem[\protect\citeauthoryear{Buchert}{Buchert}{2001}]{buchertAveragePropertiesInhomogeneous2001}
Buchert T.,  2001, \mn@doi [General Relativity and Gravitation]
  {10.1023/A:1012061725841}, 33, 1381

\bibitem[\protect\citeauthoryear{Buchert, Coley, Kleinert, Roukema  \&
  Wiltshire}{Buchert et~al.}{2016}]{buchertObservationalChallengesStandard2016}
Buchert T.,  Coley A.~A.,  Kleinert H.,  Roukema B.~F.,   Wiltshire D.~L.,
  2016, \mn@doi [International Journal of Modern Physics D]
  {10.1142/S021827181630007X}, 25, 1630007

\bibitem[\protect\citeauthoryear{Cao, Biesiada, Gavazzi, Pi{\'o}rkowska  \&
  Zhu}{Cao et~al.}{2015}]{caoCosmologyStrongLensing2015}
Cao S.,  Biesiada M.,  Gavazzi R.,  Pi{\'o}rkowska A.,   Zhu Z.-H.,  2015,
  \mn@doi [Astrophysical Journal] {10.1088/0004-637X/806/2/185}, 806, 185

\bibitem[\protect\citeauthoryear{Cappellari et~al.,}{Cappellari
  et~al.}{2006}]{cappellariSAURONProjectIV2006}
Cappellari M.,  et~al., 2006, \mn@doi [Monthly Notices of the Royal
  Astronomical Society] {10.1111/j.1365-2966.2005.09981.x}, 366, 1126

\bibitem[\protect\citeauthoryear{Cardone, Piedipalumbo  \& Scudellaro}{Cardone
  et~al.}{2016}]{cardoneCosmologicalParametersLenses2016}
Cardone V.~F.,  Piedipalumbo E.,   Scudellaro P.,  2016, \mn@doi [Monthly
  Notices of the Royal Astronomical Society] {10.1093/mnras/stv2200}, 455, 831

\bibitem[\protect\citeauthoryear{Chen, Li, Shu  \& Cao}{Chen
  et~al.}{2019}]{chenAssessingEffectLens2019}
Chen Y.,  Li R.,  Shu Y.,   Cao X.,  2019, \mn@doi [Monthly Notices of the
  Royal Astronomical Society] {10.1093/mnras/stz1902}, 488, 3745

\bibitem[\protect\citeauthoryear{Collett}{Collett}{2015}]{collettPOPULATIONGALAXYGALAXY2015}
Collett T.~E.,  2015, \mn@doi [Astrophysical Journal]
  {10.1088/0004-637X/811/1/20}, 811, 20

\bibitem[\protect\citeauthoryear{Dam, Heinesen  \& Wiltshire}{Dam
  et~al.}{2017}]{damApparentCosmicAcceleration2017}
Dam L.~H.,  Heinesen A.,   Wiltshire D.~L.,  2017, \mn@doi [Monthly Notices of
  the Royal Astronomical Society] {10.1093/mnras/stx1858}, 472, 835

\bibitem[\protect\citeauthoryear{Delubac et~al.,}{Delubac
  et~al.}{2015}]{delubacBaryonAcousticOscillations2015}
Delubac T.,  et~al., 2015, \mn@doi [Astronomy \& Astrophysics]
  {10.1051/0004-6361/201423969}, 574, A59

\bibitem[\protect\citeauthoryear{{Duley}, {Nazer}  \& {Wiltshire}}{{Duley}
  et~al.}{2013}]{DuleyTimescapeRadiation2013}
{Duley} J. A.~G.,  {Nazer} M.~A.,   {Wiltshire} D.~L.,  2013, \mn@doi
  [Classical and Quantum Gravity] {10.1088/0264-9381/30/17/175006}, \href
  {https://ui.adsabs.harvard.edu/abs/2013CQGra..30q5006D} {30, 175006}

\bibitem[\protect\citeauthoryear{Eisenstein et~al.,}{Eisenstein
  et~al.}{2005}]{eisensteinDetectionBaryonAcoustic2005}
Eisenstein D.~J.,  et~al., 2005, \mn@doi [Astrophysical Journal]
  {10.1086/466512}, 633, 560

\bibitem[\protect\citeauthoryear{Ishibashi \& Wald}{Ishibashi \&
  Wald}{2006}]{IshibashiWald2006}
Ishibashi A.,  Wald R.~M.,  2006, \mn@doi [Class. Quant. Grav.]
  {10.1088/0264-9381/23/1/012}, 23, 235

\bibitem[\protect\citeauthoryear{Jorgensen, Franx  \& Kjaergaard}{Jorgensen
  et~al.}{1995}]{jorgensenSpectroscopyS0Galaxies1995}
Jorgensen I.,  Franx M.,   Kjaergaard P.,  1995, \mn@doi [Monthly Notices of
  the Royal Astronomical Society] {10.1093/mnras/276.4.1341}, 276, 1341

\bibitem[\protect\citeauthoryear{Kass \& Raftery}{Kass \& Raftery}{1995}]{kr}
Kass R.~E.,  Raftery A.~E.,  1995, \mn@doi [J. Am. Statist. Assoc.]
  {10.1080/01621459.1995.10476572}, 90, 773

\bibitem[\protect\citeauthoryear{Koopmans}{Koopmans}{2006}]{koopmansGravitationalLensingStellar2006}
Koopmans L. V.~E.,  2006, \mn@doi [EAS Publications Series]
  {10.1051/eas:2006064}, 20, 161

\bibitem[\protect\citeauthoryear{Koopmans \& Treu}{Koopmans \&
  Treu}{2002}]{koopmansStellarVelocityDispersion2002}
Koopmans L. V.~E.,  Treu T.,  2002, \mn@doi [Astrophysical Journal]
  {10.1086/340143}, 568, L5

\bibitem[\protect\citeauthoryear{Koopmans \& Treu}{Koopmans \&
  Treu}{2003}]{koopmansStructureDynamicsLuminous2003}
Koopmans L. V.~E.,  Treu T.,  2003, \mn@doi [Astrophysical Journal]
  {10.1086/345423}, 583, 606

\bibitem[\protect\citeauthoryear{{Lane}, {Seifert}, {Ridden-Harper}  \&
  {Wiltshire}}{{Lane} et~al.}{2023}]{LaneCosmologicalFoundations2023}
{Lane} Z.~G.,  {Seifert} A.,  {Ridden-Harper} R.,   {Wiltshire} D.~L.,  2023,
  Cosmological foundations revisited with Pantheon+ (\mn@eprint {arXiv}
  {2311.01438})

\bibitem[\protect\citeauthoryear{Larena, Alimi, Buchert, Kunz  \&
  Corasaniti}{Larena et~al.}{2009}]{larenaTestingBackreactionEffects2009}
Larena J.,  Alimi J.-M.,  Buchert T.,  Kunz M.,   Corasaniti P.-S.,  2009,
  \mn@doi [Physical Review D] {10.1103/PhysRevD.79.083011}, 79, 083011

\bibitem[\protect\citeauthoryear{Leith, Ng  \& Wiltshire}{Leith
  et~al.}{2007}]{Leith_2007}
Leith B.~M.,  Ng S. C.~C.,   Wiltshire D.~L.,  2007, \mn@doi [Astrophysical
  Journal] {10.1086/527034}, 672, L91

\bibitem[\protect\citeauthoryear{Liao, Li, Wang  \& Fan}{Liao
  et~al.}{2017}]{liaoTestFLRWMetric2017}
Liao K.,  Li Z.,  Wang G.-J.,   Fan X.-L.,  2017, \mn@doi [Astrophysical
  Journal] {10.3847/1538-4357/aa697e}, 839, 70

\bibitem[\protect\citeauthoryear{Mehlert, Thomas, Saglia, Bender  \&
  Wegner}{Mehlert et~al.}{2003}]{mehlertSpatiallyResolvedSpectroscopy2003}
Mehlert D.,  Thomas D.,  Saglia R.~P.,  Bender R.,   Wegner G.,  2003, \mn@doi
  [Astronomy \& Astrophysics] {10.1051/0004-6361:20030886}, 407, 423

\bibitem[\protect\citeauthoryear{Millon et~al.,}{Millon
  et~al.}{2020}]{millonTDCOSMOExplorationSystematic2020}
Millon M.,  et~al., 2020, \mn@doi [Astronomy \& Astrophysics]
  {10.1051/0004-6361/201937351}, 639, A101

\bibitem[\protect\citeauthoryear{Navarro, Frenk  \& White}{Navarro
  et~al.}{1996}]{navarroStructureColdDark1996}
Navarro J.~F.,  Frenk C.~S.,   White S. D.~M.,  1996, \mn@doi [Astrophysical
  Journal] {10.1086/177173}, 462, 563

\bibitem[\protect\citeauthoryear{Peebles}{Peebles}{2022}]{peeblesAnomaliesPhysicalCosmology2022}
Peebles P. J.~E.,  2022, \mn@doi [Annals of Physics]
  {10.1016/j.aop.2022.169159}, 447, 169159

\bibitem[\protect\citeauthoryear{Qi, Cao, Biesiada, Zheng, Ding  \& Zhu}{Qi
  et~al.}{2019}]{qiStronglyGravitationallyLensed2019}
Qi J.,  Cao S.,  Biesiada M.,  Zheng X.,  Ding X.,   Zhu Z.-H.,  2019, \mn@doi
  [Physical Review D] {10.1103/PhysRevD.100.023530}, 100, 023530

\bibitem[\protect\citeauthoryear{R\"as\"anen, Bolejko  \&
  Finoguenov}{R\"as\"anen et~al.}{2015}]{rasanenNewTestFLRW2015}
R\"as\"anen S.,  Bolejko K.,   Finoguenov A.,  2015, \mn@doi [Physical Review
  Letters] {10.1103/PhysRevLett.115.101301}, 115, 101301

\bibitem[\protect\citeauthoryear{Refsdal}{Refsdal}{1964}]{refsdalPossibilityDeterminingHubble1964}
Refsdal S.,  1964, \mn@doi [Monthly Notices of the Royal Astronomical Society]
  {10.1093/mnras/128.4.307}, 128, 307

\bibitem[\protect\citeauthoryear{{Riess} et~al.,}{{Riess}
  et~al.}{2023}]{riessCrowdedNoMore2023}
{Riess} A.~G.,  et~al., 2023, \mn@doi [Astrophysical Journal Letters]
  {10.3847/2041-8213/acf769}, \href
  {https://ui.adsabs.harvard.edu/abs/2023ApJ...956L..18R} {956, L18}

\bibitem[\protect\citeauthoryear{Ruff, Gavazzi, Marshall, Treu, Auger  \&
  Brault}{Ruff et~al.}{2011}]{ruffSL2SGalaxyscaleLens2011}
Ruff A.~J.,  Gavazzi R.,  Marshall P.~J.,  Treu T.,  Auger M.~W.,   Brault F.,
  2011, \mn@doi [Astrophysical Journal] {10.1088/0004-637X/727/2/96}, 727, 96

\bibitem[\protect\citeauthoryear{Scrimgeour et~al.,}{Scrimgeour
  et~al.}{2012}]{Scrimgeour_2012}
Scrimgeour M.~I.,  et~al., 2012, \mn@doi [Monthly Notices of the Royal
  Astronomical Society] {10.1111/j.1365-2966.2012.21402.x}, 425, 116

\bibitem[\protect\citeauthoryear{Shu et~al.,}{Shu
  et~al.}{2015}]{shuSLOANLENSACS2015}
Shu Y.,  et~al., 2015, \mn@doi [Astrophysical Journal]
  {10.1088/0004-637X/803/2/71}, 803, 71

\bibitem[\protect\citeauthoryear{Shu et~al.,}{Shu
  et~al.}{2016a}]{shuBOSSEmissionLineLens2016}
Shu Y.,  et~al., 2016a, \mn@doi [Astrophysical Journal]
  {10.3847/0004-637X/824/2/86}, 824, 86

\bibitem[\protect\citeauthoryear{Shu et~al.,}{Shu
  et~al.}{2016b}]{shuBOSSEmissionLineLens2016a}
Shu Y.,  et~al., 2016b, \mn@doi [Astrophysical Journal]
  {10.3847/1538-4357/833/2/264}, 833, 264

\bibitem[\protect\citeauthoryear{Shu et~al.,}{Shu
  et~al.}{2017}]{shuSloanLensACS2017}
Shu Y.,  et~al., 2017, \mn@doi [Astrophysical Journal]
  {10.3847/1538-4357/aa9794}, 851, 48

\bibitem[\protect\citeauthoryear{Smale \& Wiltshire}{Smale \&
  Wiltshire}{2011}]{smaleSupernovaTestsTimescape2011}
Smale P.~R.,  Wiltshire D.~L.,  2011, \mn@doi [Monthly Notices of the Royal
  Astronomical Society] {10.1111/j.1365-2966.2010.18142.x}, 413, 367

\bibitem[\protect\citeauthoryear{Sonnenfeld, Gavazzi, Suyu, Treu  \&
  Marshall}{Sonnenfeld et~al.}{2013a}]{sonnenfeldSL2SGalaxyscaleLens2013}
Sonnenfeld A.,  Gavazzi R.,  Suyu S.~H.,  Treu T.,   Marshall P.~J.,  2013a,
  \mn@doi [Astrophysical Journal] {10.1088/0004-637X/777/2/97}, 777, 97

\bibitem[\protect\citeauthoryear{Sonnenfeld, Treu, Gavazzi, Suyu, Marshall,
  Auger  \& Nipoti}{Sonnenfeld
  et~al.}{2013b}]{sonnenfeldSL2SGalaxyscaleLens2013a}
Sonnenfeld A.,  Treu T.,  Gavazzi R.,  Suyu S.~H.,  Marshall P.~J.,  Auger
  M.~W.,   Nipoti C.,  2013b, \mn@doi [Astrophysical Journal]
  {10.1088/0004-637X/777/2/98}, 777, 98

\bibitem[\protect\citeauthoryear{Sonnenfeld, Treu, Marshall, Suyu, Gavazzi,
  Auger  \& Nipoti}{Sonnenfeld
  et~al.}{2015}]{sonnenfeldSL2SGalaxyscaleLens2015}
Sonnenfeld A.,  Treu T.,  Marshall P.~J.,  Suyu S.~H.,  Gavazzi R.,  Auger
  M.~W.,   Nipoti C.,  2015, \mn@doi [Astrophysical Journal]
  {10.1088/0004-637X/800/2/94}, 800, 94

\bibitem[\protect\citeauthoryear{{Sylos Labini}, {Vasilyev}, {Pietronero}  \&
  {Baryshev}}{{Sylos Labini} et~al.}{2009}]{Sylos_Labini_2009}
{Sylos Labini} F.,  {Vasilyev} N.~L.,  {Pietronero} L.,   {Baryshev} Y.~V.,
  2009, \mn@doi [EPL (Europhysics Letters)] {10.1209/0295-5075/86/49001}, \href
  {https://ui.adsabs.harvard.edu/abs/2009EL.....8649001S} {86, 49001}

\bibitem[\protect\citeauthoryear{Treu \& Koopmans}{Treu \&
  Koopmans}{2002}]{treuInternalStructureFormation2002}
Treu T.,  Koopmans L.,  2002, \mn@doi [Astrophysical Journal] {10.1086/341216},
  575, 87

\bibitem[\protect\citeauthoryear{Treu \& Koopmans}{Treu \&
  Koopmans}{2004}]{treuMassiveDarkmatterHalos2004}
Treu T.,  Koopmans L. V.~E.,  2004, \mn@doi [Astrophysical Journal]
  {10.1086/422245}, 611, 739

\bibitem[\protect\citeauthoryear{{Trotta}}{{Trotta}}{2007}]{t07}
{Trotta} R.,  2007, \mn@doi [\mnras] {10.1111/j.1365-2966.2007.11738.x}, \href
  {https://ui.adsabs.harvard.edu/abs/2007MNRAS.378...72T} {378, 72}

\bibitem[\protect\citeauthoryear{Trotta}{Trotta}{2017}]{trottaBayesianMethodsCosmology2017}
Trotta R.,  2017, Bayesian {{Methods}} in {{Cosmology}} (\mn@eprint {arxiv}
  {1701.01467}), \mn@doi{10.48550/arXiv.1701.01467}

\bibitem[\protect\citeauthoryear{Wagner}{Wagner}{2020}]{wagnerSelfgravitatingDarkMatter2020}
Wagner J.,  2020, \mn@doi [International Journal of Modern Physics D]
  {10.1142/S0218271820430178}, 29, 2043017

\bibitem[\protect\citeauthoryear{{Wiltshire}}{{Wiltshire}}{2007a}]{wiltshireCosmicClocks2007a}
{Wiltshire} D.~L.,  2007a, \mn@doi [New Journal of Physics]
  {10.1088/1367-2630/9/10/377}, \href
  {https://ui.adsabs.harvard.edu/abs/2007NJPh....9..377W} {9, 377}

\bibitem[\protect\citeauthoryear{{Wiltshire}}{{Wiltshire}}{2007b}]{wiltshireExactSolution2007b}
{Wiltshire} D.~L.,  2007b, \mn@doi [Physical Review Letters]
  {10.1103/PhysRevLett.99.251101}, \href
  {https://ui.adsabs.harvard.edu/abs/2007PhRvL..99y1101W} {99, 251101}

\bibitem[\protect\citeauthoryear{Wiltshire}{Wiltshire}{2009a}]{wiltshireTimeTimescapeEinstein2009a}
Wiltshire D.~L.,  2009a, \mn@doi [International Journal of Modern Physics D]
  {10.1142/S0218271809016193}, 18, 2121

\bibitem[\protect\citeauthoryear{Wiltshire}{Wiltshire}{2009b}]{wiltshireAverageObservationalQuantities2009}
Wiltshire D.~L.,  2009b, \mn@doi [Physical Review D]
  {10.1103/PhysRevD.80.123512}, 80, 123512

\bibitem[\protect\citeauthoryear{Wiltshire}{Wiltshire}{2014}]{Wiltshire_2014_cosmic}
Wiltshire D.~L.,  2014, in {Perez Bergliaffa} S.,  {Novello} M.,  eds, ,
  {Proceedings of the XVth Brazilian School of Cosmology and Gravitation}.
{Cambridge Scientific Publishers}, pp 203--244 (\mn@eprint {arXiv} {1311.3787})

\bibitem[\protect\citeauthoryear{Wong et~al.,}{Wong
  et~al.}{2020}]{wongH0LiCOWXIIIMeasurement2020}
Wong K.~C.,  et~al., 2020, \mn@doi [Monthly Notices of the Royal Astronomical
  Society] {10.1093/mnras/stz3094}, 498, 1420

\makeatother
\end{thebibliography}
\newpage
\appendix
\section{Priors}
The priors used for the MCMC sampling procedure and Bayesian analysis are listed in Table \ref{tab:MCMC_priors} and Table \ref{tab:Bayes_Priors}.
\begin{table*}
\begin{tabular}{ |p{2cm}||p{2cm}|p{2cm}|p{2cm}|p{2cm}|p{2cm}|p{2cm}| }
\hline
\multicolumn{7}{|c|}{MCMC Priors} \\
\hline
Lens Model& $\gamma$ &$\gamma_z$& $\gamma_s$ & $\beta$ &$\delta$ &$f_{\rm v0}$/$\Omega_{\rm M0}$\\
\hline
EPL (full)  & (1.0, 3.0) & (-1, 1) & (0,1) & $\mathcal{N}$(-0.08, 0.44) & (1.8, 2.6) & (0, 1)\\
EPL ($\gamma = \gamma_0$)  & (1.8, 2.2) & --  & -- & $\mathcal{N}$(-0.08, 0.44) & (1.8, 2.6) & (0, 1) \\
SPL& (1.8, 2.2) & -- & -- & -- & -- & (0, 1) \\
SIS & -- & -- &  -- & -- & -- & (0, 1) \\
\end{tabular}
{\caption{Priors assumed for MCMC sampling. In general wide uniform priors are taken, with the except of the stellar orbital anisotropy parameter $\beta$, which is assumed to have a Gaussian distribution. \label{tab:MCMC_priors} }}
\end{table*}

\begin{table*}
\begin{tabular}{ |p{0.28\columnwidth}||p{0.28\columnwidth}|p{0.28\columnwidth}| p{0.28\columnwidth}|p{0.28\columnwidth}||p{0.28\columnwidth}| }
\hline
\multicolumn{6}{|c|}{Bayes Factors Priors} \\
\hline
Lens Model& $\gamma$ & $\delta$ & $\beta$ & $f_{\rm v0}$ & $\Omega_{\rm M0}$\\
\hline
EPL   & (1.2, 2.8) & (2.003, 2.343)  & $\mathcal{N}$(-0.08, 0.44) & (0.5, 0.799) & (0.143, 0.487) \\
SPL& (1.2, 2.8)  & --  & -- & (0.5, 0.799) & (0.143, 0.487) \\
SIS & -- & --  & -- & (0.5, 0.799) & (0.143, 0.487) \\
\end{tabular}
\caption{{Priors used for establishing Bayes factors between cosmological models. }
\label{tab:Bayes_Priors} }
\end{table*}

\section{Model derivation and explanation of parameters}
\label{EPL Appendix}

In this section we will discuss the various assumptions that go into the extended power-law model of \citet{koopmansGravitationalLensingStellar2006} for elliptical galaxies as gravitational lenses. The extended power-law model, as well as variants SPL and SIS, are derived from the radial Jeans equation here given in spherical coordinates $(r,\theta,\phi)$: 
\begin{equation}
    \frac{\dd}{\dd r} [\nu(r) \, \sigma^2_r] + \frac{2\beta}{r} \, \nu(r) \, \sigma^2_r = -\nu(r) \frac{\dd \Phi}{\dd r} \, ,
\end{equation}
where 
\begin{equation}
    \frac{\dd \Phi}{\dd r} = \frac{GM(r)}{r^2}  
\end{equation}
Here, the velocity dispersion $\sigma_r$ is defined as follows with $f$ being the distribution function of the stars within the galaxy. 
\begin{equation}
    \sigma_r^2 = \frac{1}{\nu} \int v_r f \dd^3v
\end{equation}
The Jeans equation looks at the average motion of the distribution of stars defined by the luminous matter density distribution $\nu(r)$ within a Newtonian gravitational potential created by the total matter density $\rho(r)$, including dark matter. The parameter $\beta(r)$ denotes the anisotropy of the stellar velocity dispersion and is also known as the stellar orbital anisotropy. These are parametrised as such for the extended power-law,
\begin{align}
    \rho (r) = \rho\Z0 (r/r\Z0)^{-\gamma} \\
    \nu (r) = \nu\Z0 (r/r\Z0)^{-\delta} \\
    \beta (r) = 1 - \sigma^2_\theta / \sigma^2_r \, ,
\end{align}
where $\sigma_\theta$ and $\sigma_r$ are the tangential and radial velocity dispersions respectively. With $\beta(r) \neq 0$, we can account the average orbits of stars deviating from circular paths, however the total mass density will remain spherical - it does not disrupt the spherical dark matter halo of the galaxy.

In order to solve the Jeans equation, numerous assumptions are made:
\begin{itemize}
    \item the system is static i.e., $\partial_t  = 0$. This feels like a reasonable assumption to make as the structure of the elliptical galaxy should not be expected to change over the course of lensing measurements. 
    \item the stress tensor $\sigma_{ij}$ is diagonal, such that $\expval{v_i v_j} = 0$ if $i \neq j$.
    \item spherical symmetry applies such that $\sigma_{r \phi}^2 = \sigma_{r \theta}^2 = 0$ and $\rho$ and $\nu$ only depend on $r$ with no angular component.
    \item collisionless matter. The Jeans equation is originally derived from the collisionless Boltzman equation and as such does not permit interactions between the constituent stars within galaxies. 
\end{itemize}
After making these assumptions, the radial Jeans equation can be solved for the radial velocity dispersion in terms of the luminous matter distribution and the total mass inside a sphere of radius $r$,
\begin{equation}
    \sigma^2_r (r) = \frac{G \int^\infty_r \dd r' {r'}^{2\beta -2} \, \nu(r') M(r') }{r^{2\beta} \, \nu(r)} \,.
\end{equation}
We can then define the mass contained within a cylinder of radius $R_E$, the Einstein radius, which quantifies the mass of the lens galaxy,
\begin{equation}
    M_E = \int^{R_E}_0 \dd R\,2\pi R' \Sigma(R') \, ,
\end{equation}
where  $\Sigma(R)$ is the mass density projected into the lens plane,
\begin{equation}
\Sigma(R) = \int^\infty_{-\infty} \dd Z\, \rho(r) = \int^\infty_{-\infty} \dd Z\, \rho\Z0\, r_0^\gamma (Z^2 + R^2)^{-\gamma/2} \,,
\end{equation}
\begin{equation}
    \Sigma(R) = \sqrt{\pi} \,  R^{1-\gamma} \frac{\Gamma((\gamma -1)/2)}{\Gamma(\gamma/2)} \rho\Z0 \, r_0^\gamma \,.
\end{equation}
Here $R$ is the radius of the galaxy in the lens plane and $Z$ is the distance along the line of sight perpendicular to the lens plane, $r$ is the spherical radius as previous such that $r^2 = R^2 + Z^2$. 

The mass contained inside the Einstein radius is therefore, 
\begin{equation}
    M_E = 2\pi^{3/2} \, \frac{R_E^{3-\gamma}}{3-\gamma} \frac{\Gamma((\gamma -1)/2)}{\Gamma(\gamma/2)} \rho\Z0 \, r_0^\gamma \,.
\end{equation}
The total mass within a sphere of radius $r$ can also be calculated,
\begin{equation}
    M(r) = \int^r_0 \dd r' 4\pi \, {r'}^2 \, \rho(r') = 4\pi \, \rho\Z0 \,  r_0^\gamma \frac{r^{3-\gamma}}{3-\gamma}
\end{equation}
which can be written in terms of the mass $M_E$ contained within the cyclinder as
\begin{equation}
    \sigma^2_r (r) = \frac{2}{\sqrt{\pi}} \frac{GM_E}{R_E} \frac{1}{\xi - 2\beta} \frac{\Gamma((\gamma -1)/2)}{\Gamma(\gamma/2)} \left( \frac{r}{R_E} \right)^{2-\gamma}
\end{equation}
where $\xi = \gamma + \delta - 2$. The velocity dispersion from observation is the component of the luminosity weighted average along the line of sight and pver the effective spectroscopic aperture $R_A$. This can be written as:
\begin{equation}
    \sigma^2_\parallel (\leq R_A) = \frac{\int^{R_A}_0 \int^\infty_{-\infty} \dd R \, \dd Z \, 2\pi R \, \sigma^2_r (r) \left(1-\beta \frac{R^2}{r^2}\right) \nu(r) } {\int^{R_A}_0 \int^\infty_{-\infty} \dd R \, \dd Z \, 2\pi R \nu(r)} \,.
\end{equation}
By substituting in the expression for $\sigma^2_r$ and the power-law of $\nu(r)$ we arrive at 
\begin{multline}
     \sigma^2_\parallel (\leq R_A) =  \frac{2}{\sqrt{\pi}} \frac{GM_E}{R_E} \frac{3-\delta}{(\xi -2\beta)\,(3-\xi)} \left[ \frac{\Gamma \left(\frac{\xi-1}{2}\right)}{\Gamma \left(\frac{\xi}{2}\right)} - \beta \frac{\Gamma \left(\frac{\xi+1}{2}\right)}{\Gamma \left(\frac{\xi+2}{2}\right)} \right] \\ \times \frac{\Gamma \left(\frac{\gamma}{2}\right) \Gamma \left(\frac{\delta}{2}\right)}{\Gamma \left(\frac{\gamma-1}{2}\right) \Gamma \left(\frac{\delta-1}{2}\right)} \left( \frac{R_A}{R_E} \right)^{2-\gamma} \,.
\end{multline}
Using the deflecting mass found from the lensing equation
\begin{equation}
    M_E = \frac{c^2 \, \theta_E}{4G} \frac{D_s \, D_l}{D_{ls}}\,,
\end{equation}
$R_A = D_l \theta_A$ and $R_E = \theta_E D_l$ we get,
\begin{multline}
     \sigma^2_\parallel (\leq R_A) =  \frac{2}{\sqrt{\pi}} \frac{D_s \, \theta_E \, c^2}{D_{ls}} \frac{3-\delta}{(\xi -2\beta)(3-\xi)} \left[ \frac{\Gamma \left(\frac{\xi-1}{2}\right)}{\Gamma \left(\frac{\xi}{2}\right)} - \beta \frac{\Gamma \left(\frac{\xi+1}{2}\right)}{\Gamma \left(\frac{\xi+2}{2}\right)} \right] \\ \times \frac{\Gamma \left(\frac{\gamma}{2}\right) \Gamma \left(\frac{\delta}{2}\right)}{\Gamma \left(\frac{\gamma-1}{2}\right) \Gamma \left(\frac{\delta-1}{2}\right)} \left( \frac{\theta_A}{\theta_E} \right)^{2-\gamma} \,,
\end{multline}
which is the expression we call the extended power-law. The use of the lens equation also is built upon several assumptions:
\begin{itemize}
    \item the weak field limit of general relativity applies and that $\Phi/c^2\ll1$,
    \item small deflection angles,
    \item the Born approximation applies.
\end{itemize}


\bsp	
\label{lastpage}
\end{document}